\documentclass[prc]{revtex4}
\usepackage{graphicx}
\begin{document}
%\linenumbers

\newcommand\ie {{\it i.e.}}
\newcommand\eg {{\it e.g.}}
\newcommand\etc{{\it etc.}}
\newcommand\cf {{\it cf.}}
\newcommand\etal {{\it et al.}}
\newcommand{\be}{\begin{eqnarray}}
\newcommand{\ee}{\end{eqnarray}}
\newcommand{\jp}{$ J/ \psi $}
\newcommand{\pp}{$ \psi^{ \prime} $}
\newcommand{\ppp}{$ \psi^{ \prime \prime } $}
\newcommand{\dd}[2]{$ #1 \overline #2 $}
\newcommand\noi {\noindent}
%\tightenlines

\title{Contribution from Intrinsic Charm Production to Fixed-Target
Interactions with the SMOG Device at LHCb}

\author{R. Vogt}
\affiliation
{Nuclear and Chemical Sciences Division,
Lawrence Livermore National Laboratory, Livermore, CA 94551,
USA}
\affiliation
    {Department of Physics and Astronomy,
University of California, Davis, CA 95616,
USA}

\begin{abstract}
  {\bf Background:} Intrinsic charm, nonperturbative charm in the hadron
  wavefunction, has long been speculated but has never been satisfactorily
  proven.  Open charm and $J/\psi$ measurements in a fixed-taget configuration
  at the LHCb searched for this contribution but reported no evidence.
  {\bf Purpose:} $\overline D$ meson and $J/\psi$ production is calculated for
  the SMOG fixed-target configuration in the LHCb experiment using 
  a combination of perturbative QCD and intrinsic charm to see whether intrinsic
  chaarm would indeed be observable in the SMOG kinematics.
  {\bf Methods:} Open charm and $J/\psi$ production is calculated to
  next-to-leading order in perturbative QCD.   Because a gas jet nuclear
  target is used, cold nuclear matter effects
  are included in the perturbative calculations.  The
  intrinsic charm is calculated assuming production from a
  $|uud c \overline c \rangle$ Fock state.
  {\bf Results:}  The differential rapidity and transverse momentum
  distributions in $p+{\rm Ne}$, $p+{\rm He}$ and $p+{\rm Ar}$ fixed-target
  interactions
  are calculated in the SMOG acceptance and compared to data.  The
  predicted asymmetries between $\overline D$ (leading charm) and $D$
  (nonleading charm) are also shown.
  {\bf Conclusions:}  The contribution from intrinisic charm is small and
  decreases with center of mass energy.  The calculations agree well with the
  current SMOG data, with or without intrinsic charm.
\end{abstract}
\maketitle

\section{Introduction}

%Feel good motivational paragraph from previous paper.

There has been a renewed interest in intrinsic charm in the last several years
and a potential resolution to whether it exists or not may be within view.
The LHCb collaboration measured $Z + c$-jets relative to 
$Z +$jets at $\sqrt{s} = 13$~TeV in $p+p$ collisions at forward rapidity and found
that the most forward rapidity data could not be explained without an intrinsic charm 
contribution on the order of 1\% \cite{LHCb_intc}.  These data were 
employed in the recent global proton parton distribution function analysis
by the NNPDF collaboration that 
extracted charm quark distribution to NNLO \cite{NNPDF}.
Without these LHCb data, the significance of the intrinsic 
charm signal was 2.5$\sigma$ and, with it, the significance increased to
3$\sigma$ \cite{NNPDF}.  High precision data from forward charm hadron
production at fixed-target energies, along with charm 
measurements at the future electron-ion collider, could provide confirmation of
the result and increase the significance, leading to greater significance,
potentially leading to a report of discovery of intrinsic 
charm rather than only evidence of its production.

The calculations discussed in this work follow those of Ref.~\cite{RV_IC_EN}
where $J/\psi$ and $\overline D$ meson production from a combined model of perturbative
QCD and intrinsic charm were studied
over a wide energy range, from $\sqrt{s_{NN}} = 8.77$~GeV for the proposed NA60+
to the highest
energy at the LHC, $\sqrt{s} = 13$~TeV.  It was shown that, due to the boost
from the center of mass energy when converting from $x_F$ to rapidity, the
intrinsic charm contribution, while independent of
$x_F$, is strongly dependent on rapidity.  Therefore, there should be a stronger
signal from intrinsic charm at lower center of mass energies.  The rapidity
range covered by an experiment also has a significant
effect on the observed $p_T$ distribution from intrinsic charm with the
percentage of the total intrinsic charm contribution decreasing with center
of mass energy for acceptance at midrapidity. 
The larger the forward (or backward) rapidity coverage at high energies
\cite{RV_IC_EN}, the larger the potential contribution from intrinsic charm.

The SMOG device was employed in the LHCb detector to take $J/\psi$ and $D$ meson
data at fixed-target energies from 68.5 to 110.4~GeV.  These measurements are
discussed in Sec.~\ref{SMOG}.
In this work, $J/\psi$ and $D^0$ meson production by perturbative QCD is
presented in Sec.~\ref{pQCD}.  The $p+p$ distributions are shown, including
their mass and scale uncertainties. The cold nuclear matter effects employed in
the calculation are briefly introduced in Sec.~\ref{CNM}.  
The $p+p$ distributions from intrinsic charm in the SMOG kinematics are shown in
Sec.~\ref{IC}. Finally, the two contributions are combined and compared with
fixed-target data from LHCb employing the SMOG device
\cite{SMOG,SMOGpNeJ,SMOGpNeD} in Sec.~\ref{comp},
along with predictions for $D$ meson
asymmetries.  The conclusions are presented in Sec.~\ref{conclusions}.

\section{The SMOG device at LHCb}
\label{SMOG}

The LHCb detector, designed to study charm and bottom hadrons,
has a single arm spectrometer covering the forward
pseudorapidity region $2 < \eta < 5$.  It normally operates in collider mode.
However, the system for measuring overlap with gas (SMOG) device \cite{SMOG}
enables low pressure gases to be injected into the beam pipe, near their
silicon-strip vertex detector, allowing LHCb to function also as a fixed-target
experiment.  SMOG injects noble gases, He, Ne and Ar so far, which can interact
with either proton or nuclear beams circulating in the LHC.  To this point,
$p+{\rm He}$ at $\sqrt{s_{NN}} = 86.6$~GeV and $p+{\rm Ar}$ collisions 
at $\sqrt{s_{NN}} = 110.4$~GeV \cite{SMOG} and $p+{\rm Ne}$ collisions at
$\sqrt{s_{NN}} = 68.5$~GeV have been reported \cite{SMOGpNeJ,SMOGpNeD}.

Because of the boost, the LHCb acceptance covers a range from backward rapidity
to near central rapidity in the fixed-target mode, $y_{\rm cm}$ from $\sim -2.5$
to $\sim 0$.  Data were taken for proton beams of 2.55, 4 and 6.5~TeV on neon, helium and
argon gas targets respectively.  Data on $J/\psi$ and $D^0$
production were taken in all cases.  The difference in center of mass energies
for the three systems resulted in a slight difference in rapidity coverage with
$-2.29 \leq y_{\rm cm} \leq 0$ for $p+{\rm Ne}$,
$-2.53 \leq y_{\rm cm} \leq 0.07$ for $p+{\rm He}$, and
$-2.77 \leq y_{\rm cm} \leq -0.17$ for $p+{\rm Ar}$.  In all cases the
transverse momentum range probed was $p_T \leq 8$~GeV.

The spectrometer
permits the collaboration to study $J/\psi$ through their decay to muon pairs
while the silicon-strip detector makes it possible to reconstruct $D^0$ mesons
via their two-body decays to $K^- \pi^+$ (as well as their $\overline D^0$
charge conjugates).
The data reported in Ref.~\cite{SMOG} were collected under specific beam
conditions with no proton bunch crossings at the $p+p$ interaction point to
reduce background.  

The $J/\psi$ and $D^0$ production cross sections were obtained for $p+{\rm He}$
collisions because a luminosity determination was available for that system.
However, no such determination was made for the $p+{\rm Ar}$ system.  Therefore,
the yields, normalized to unity, were reported instead.  The $p+{\rm Ne}$ data 
are also reported in terms of production cross sections \cite{SMOGpNeJ,SMOGpNeD}.

The rapidity range covered in this fixed-target setup should allow coverage up
to $x \sim 0.37$ for $D^0$ mesons, permitting a test of intrinsic charm
production.  In their first paper \cite{SMOG}, the LHCb collaboration found
their data to be consistent with perturbative calculations that did not
include any intrinsic charm production.  The calculations were performed with
HELAC-ONIA \cite{Helac1,Helac2,Helac3}.  The only cold nuclear matter effect
included was the modification of the parton distributions in nuclei, implemented
using the nCTEQ15 nuclear parton densities (nPDFs) \cite{nCTEQ15}.

The rapidity and $p_T$ distributions have been reported for $J/\psi$ and
$D^0/\overline D^0$ production for all systems studied
\cite{SMOG,SMOGpNeJ,SMOGpNeD}.  In addition, the
asymmetry between $D^0$ and $\overline D^0$ mesons has been reported for
$p+{\rm Ne}$ collisions \cite{SMOGpNeD}.

\section{$J/\psi$ and $\overline D^0$ Production in Perturbative QCD}
\label{pQCD}

The open charm and charmonium production cross sections in perturbative QCD are treated 
similarly in this work.  Because charmonium is calculated in the color evaporation model, 
the main difference is in the partonic center of mass energy range, $\hat{s}$, 
of the integration.  The open charm cross section is integrated over the full available 
energy range of $\hat{s}$ while the upper limit of the $\hat{s}$ integration range for
$J/\psi$ is the square of the $D \overline D$ mass threshold. The next-to-leading order 
(NLO) heavy flavor cross section is obtained using the HVQMNR code \cite{HVQMNR}, both 
for open charm and charmonium. 

More recently, the improved color evaporation model has been developed \cite{ICEM} 
which uses the quarkonium mass itself as the limit of integration and the quarkonium 
momentum is modified according to the quarkonium mass, originally still within the 
context of the HVQMNR code \cite{ICEM} and later to LO in $k_T$ factorization 
\cite{Cheung:2019} and
NLO in collinear factorization to study the polarization as well \cite{Cheung:202x}.  
These changes slightly modify the slope of the $J/\psi$ $p_T$ distribution and also 
result in different kinematic distributions for the $J/\psi$ and $\psi$(2S) \cite{ICEM}
although these changes are not large\cite{ICEM,Cheung:202x}.  
Open charm is discussed first, followed by quarkonium.  The cold nuclear matter effects, 
which will be applied to both open charm and charmonium production, will be discussed 
in Sec.~\ref{CNM}.

The perturbative open heavy flavor (OHF) cross section can be schematically
represented as
\be
\sigma_{\rm OHF}(pp) = \sum_{i,j} 
\int_{4m^2}^{\infty} d\hat{s}
\int dx_1 \, dx_2~ F_i^p(x_1,\mu_F^2,k_{T_1})~ F_j^p(x_2,\mu_F^2,k_{T_2})~ 
\hat\sigma_{ij}(\hat{s},\mu_F^2, \mu_R^2) \, \, , 
\label{sigOHF}
\ee
where $ij = g+g, q + \overline q$ or $q(\overline q)+ g$ and
$\hat\sigma_{ij}(\hat {s},\mu_F^2, \mu_R^2)$
is the partonic cross section for initial state $ij$ with the
$q(\overline q)+g$ process oappearing at next-to-leading order in $\alpha_s$.  
The cross section and parton distribution functions are calculated at factorization 
scale $\mu_F$ and renormalization scale $\mu_R$. The next-to-leading order heavy flavor 
cross section is obtained using the HVQMNR code \cite{HVQMNR}.
Both calculations employ the same set of values for the charm quark
mass, $m$, and scales $\mu_F$ and $\mu_R$,
determined from a fit to the total $c \overline c$
cross section at NLO in Ref.~\cite{NVF}:
$(m,\mu_F/m_T, \mu_R/m_T) = (1.27 \pm 0.09 \, {\rm GeV}, 2.1^{+2.55}_{-0.85}, 1.6^{+0.11}_{-0.12})$.
The scales are defined relative to the transverse mass of the pair,
$\mu_{F,R} \propto m_T = \sqrt{m^2 + p_T^2}$ where 
the $p_T$ is the $c \overline c$ pair $p_T$, 
$p_T^2 = 0.5(p_{T_Q}^2 + p_{T_{\overline Q}}^2)$.  

The charm quarks become $D$ mesons by applying a fragmentation function, $D(z)$.  
The default fragmentation function in HVQMNR, applied to open heavy flavor
production only, is the Peterson function \cite{Pete},
\begin{eqnarray}
  D(z) = \frac{z(1-z)^2}{((1-z)^2 + z \epsilon_P)^2} \, \, ,
  \label{Eq.Pfun}
\end{eqnarray}
where $z$ represents the fraction of the parent heavy
flavor quark momentum carried by the resulting heavy flavor hadron.
The default parameter $\epsilon_P$, $\epsilon_P = 0.06$, employed in HVQMNR 
produces softer charm meson $p_T$
distributions than supported by data, even when intrinsic parton transverse
momentum is included \cite{MLM2},  In Ref.~\cite{RV_azi1} the value of
$\epsilon_P$ was decreased to match the FONLL $D$ meson $p_T$
distributions with intrinsic $k_T$ included.  
The same procedure is followed here with the same value of
$\epsilon_P$, 0.008 \cite{RV_azi1}.

Note that there is some correlation in the choices of $\epsilon_P$ and
$\langle k_T^2 \rangle$.  A larger $\epsilon_P$, closer to the default value of
0.06 in the HVQMNR code, results in a more steeply falling $p_T$ distribution,
the effect of which could be reversed by a sufficiently large
$\langle k_T^2 \rangle$, as discussed in Ref.~\cite{MLM2}.  However, a strong
correlation only exists at low $\sqrt{s_{NN}}$ when the average $p_T$ of the
heavy quark is low and $\langle p_T^2 \rangle \approx \langle k_T^2 \rangle$.
Above $\sqrt{s_{NN}}$ of a few tens of GeV,
$\langle p_T^2 \rangle > \langle k_T^2 \rangle$ and the effect of $k_T$
broadening on the $p_T$ distribution is decreased, even though
$\langle k_T^2 \rangle$ growns slowly with $\sqrt{s_{NN}}$.  On the other hand,
a fixed $\epsilon_P$ reduces the charm quark momentum by the same fraction at
all energies.  Thus increasing $\sqrt{s_{NN}}$ weakens the correlation between
$\epsilon_P$ and $\langle k_T^2 \rangle$.  See Ref.~\cite{RV_azi1} for
comparisons of $D$ and $B$ meson distributions with different choices of
$\epsilon_P$ and $\langle k_T^2 \rangle$ compared to FONLL at
$\sqrt{s} = 7$~TeV.

The parton densities in Eq.~(\ref{sigOHF}) include intrinsic $k_T$, required
to keep the pair cross section
finite as $p_{T_{Q \overline Q}} \rightarrow 0$.
They are assumed to factorize into the normal collinear factorization
parton densities and a $k_T$-dependent function,
\be
F^p(x,\mu_F^2,k_T) = f^p(x,\mu_F^2)G_p(k_T) \, \, . \label{PDFfact}
\ee
The CT10 proton parton distribution functions (PDFs)
\cite{CT10} are employed in the calculations of $f^p(x,\mu_F^2)$.

Results on open heavy flavors at fixed-target energies 
indicated that some level of
transverse momentum broadening was needed to obtain agreement with the
fixed-target data once fragmentation was included \cite{MLM1}.
Broadening has typically been modeled by intrinsic transverse
momentum, $k_T$, added to the parton densities and playing the role of
low transverse momentum QCD resummation \cite{CYLO}.  

In the HVQMNR code, an intrinsic $k_T$ is added each final state charm quark, 
rather than to the initial state, as in the case of
Drell-Yan production \cite{CYLO}. 
In the initial-state, the intrinsic $k_T$ function multiplies the parton
distribution functions for both hadrons, 
assuming the $x$ and $k_T$ dependencies factorize,
as in Eq.~(\ref{PDFfact}).  
At leading order, there is no difference between an initial (on the partons) or 
final-state (on the produced charm quarks) $k_T$ kick.  
However, at NLO, when there is a
light parton in the final state, the correspondence can be inexact.  
The difference between the two implementations is small if
$\langle k_T^2 \rangle \leq 2-3$ GeV$^2$ \cite{MLM1}.
The rapidity distributions are independent of the intrinsic $k_T$.

If the $k_T$ kick is not too large, it does not matter whether it 
is added in the initial or final state.
A Gaussian distribution is employed for
$G_p(k_T)$ in Eq.~(\ref{PDFfact}) \cite{MLM1}, 
\begin{eqnarray}
G_p(k_T) = \frac{1}{\pi \langle k_T^2 \rangle_p} \exp(-k_T^2/\langle k_T^2
\rangle_p) \, \, .
\label{intkt}
\end{eqnarray}
The effect of the $k_T$ kick alone hardens
the single charm meson $p_T$ distribution.
In Ref.~\cite{MLM1}, $\langle k_T^2 \rangle_p = 1$ GeV$^2$, in combination with
fragmentation using the default Peterson parameter $\epsilon_P$, was chosen
to describe low energy fixed-target charm production.  

The broadening is applied by boosting the transverse momentum
of the $c \overline c$ pair
(plus light parton at NLO)
to its rest frame
from the longitudinal center-of-mass frame.
The transverse momenta of the incident partons, $\vec k_{T_1}$ and
$\vec k_{T_2}$, or, in this case, the final-state $c$ and $\overline c$, are
redistributed isotropically with unit modulus, according to
Eq.~(\ref{intkt}), preserving momentum conservation.
Once boosted back to the initial frame, transverse momentum of
the $c \overline c$ pair changes from $\vec p_{T_{Q \overline Q}}$ to
$\vec p_{T_{Q \overline Q}} + \vec k_{T 1} + \vec k_{T 2}$ \cite{MLM2}.  

The broadening effect 
decreases as $\sqrt{s}$ increases because the perturbatively-calculated average 
$p_T$ of the $c \overline c$ pair also increases 
with $\sqrt{s}$.  The value of $\langle k_T^2 \rangle_p$ is assumed to
increase with $\sqrt{s}$ so that effect is non-negligible for low $p_T$ 
heavy flavor production at higher energies.
The energy dependence of $\langle k_T^2 \rangle$ in Ref.~\cite{NVF} is
\begin{eqnarray}
  \langle k_T^2 \rangle_p = \left[ 1 + \frac{1}{n} \ln
    \left(\frac{\sqrt{s} ({\rm GeV})}{20 \,
    {\rm GeV}} \right) \right] \, \, {\rm GeV}^2 \, \, 
\label{eq:avekt}
\end{eqnarray}
with $n = 12$ for $J/\psi$ production \cite{NVF} so that $\langle k_T^2 \rangle$ increases
slowly with energy.  At SMOG energies, 
$\langle k_T^2 \rangle = 1.10$, 1.12, and 1.14~GeV$^2$ for $\sqrt{s} = 68.5$,
86.6 and 110.4~GeV respectively.  
The values of $\langle k_T^2 \rangle$ are thus 
well below the limit of applicability proposed in Ref.~\cite{MLM1}.

The model calculation of $J/\psi$ production is now described.
The $J/\psi$ production mechanism remains an unsettled question, with a number
of approaches having been introduced \cite{HPC,NRQCD,ICEM}.  In the calculations
presented here, the Color Evaporation Model \cite{HPC} is employed.  This model,
together with the Improved Color Evaporation Model \cite{ICEM}, can describe
the $J/\psi$ rapidity and transverse momentum distributions, including at low
$p_T$ where other approaches have some difficulties and may require a $p_T$
cut \cite{QWG_rev}.  

The CEM assumes that some fraction, $F_C$, of the $c \overline c$ pairs
produced with a pair mass below the $D \overline D$ pair mass threshold
will go on mass shell as a $J/\psi$,
\be
\sigma_{\rm CEM}(pp) = F_C \sum_{i,j} 
\int_{4m^2}^{4m_H^2} d\hat{s}
\int dx_1 \, dx_2~ F_i^p(x_1,\mu_F^2,k_{T_1})~ F_j^p(x_2,\mu_F^2,k_{T_2})~ 
\hat\sigma_{ij}(\hat{s},\mu_F^2, \mu_R^2) \, \, .
\label{sigCEM}
\ee
The same mass and scale parameters are employed as in Eq.~(\ref{sigOHF}).
However, now an upper limit of $4m_D^2$ is applied and the normalization factor
$F_C$ is obtained by fitting the energy dependence of the $J/\psi$ forward
cross section \cite{NVF}.

At LO in the CEM,
the $J/\psi$ $p_T$, $p_{T_{Q \overline Q}}$ above,
is zero.  Thus $k_T$ broadening is required to keep the
$p_T$ distribution finite as $p_T \rightarrow 0$.  The intrinsic $k_T$
broadening in $J/\psi$ production is handled the same way as for open heavy
flavor production, outlined above.  However, in this case, $D(z)$ is not applied, hadronization is implied by the factor
$F_C$.  (Note that, for simplicity, in the rest of this paper,
when open charm meson distributions are
presented, $p_T$ refers to the single charm hadron transverse momentum
distribution while, when $J/\psi$ distributions are discussed, $p_T$ refers
to the transverse momentum distribution of the $J/\psi$.)

The mass and scale uncertainties on the $p+p$ 
cross sections are calculated from the one standard deviation uncertainties on
$(m,\mu_F/m_T, \mu_R/m_T)$.  If the central, upper and lower limits
of $\mu_{R,F}/m_T$ are denoted as $C$, $H$, and $L$ respectively, the seven
sets of scale values used to calculate the uncertainty are  $\{(\mu_F/m_T,\mu_F/m_T)\}$ =
$\{$$(C,C)$, $(H,H)$, $(L,L)$, $(C,L)$, $(L,C)$, $(C,H)$, $(H,C)$$\}$.   At
each point, the set giving the highest (lowest) value of the cross section
is used to calculated the maximum (minimum) contribution to the scale
uncertainty.  The
mass uncertainty is based on the one standard deviation uncertainty on the charm quark
mass, $\pm 0.09$~GeV.  The higher mass contributes to the lower limit of the cross section while the lower mass contributes to the upper limit.
The uncertainty band can be obtained for the best fit sets
\cite{NVF} by
adding the uncertainties from the mass and scale variations in 
quadrature. The result at each point,
\begin{eqnarray}
\frac{d\sigma_{\rm max}}{dX} & = & \frac{d\sigma_{\rm cent}}{dX} 
+ \sqrt{\left(\frac{d\sigma_{\mu ,{\rm max}}}{dX} -
  \frac{d\sigma_{\rm cent}}{dX}\right)^2
  + \left(\frac{d\sigma_{m, {\rm max}}}{dX} -
  \frac{d\sigma_{\rm cent}}{dX}\right)^2} \, \, , \label{sigmax}
\\
\frac{d\sigma_{\rm min}}{dX} & = & \frac{d\sigma_{\rm cent}}{dX} 
- \sqrt{\left(\frac{d\sigma_{\mu ,{\rm min}}}{dX} -
  \frac{d\sigma_{\rm cent}}{dX}\right)^2
  + \left(\frac{d\sigma_{m, {\rm min}}}{dX} -
  \frac{d\sigma_{\rm cent}}{dX}\right)^2} \, \, , \label{sigmin}  
\end{eqnarray}
defines the uncertainty on the cross section.
The kinematic observables, denoted by $X$, are $y$ and $p_T$ in this case.
The calculation labeled ``cent'' employs the central values of
$m$, $\mu_F/m_T$ and $\mu_R/m_T$.  The calculations with subscript
$\mu$ keep the mass fixed to the central value while the scales are varied.
On the other hand, in the calculations with subscript $m$, the scales are fixed
to their central values while the mass is varied
between its upper and lower limits.  

\begin{figure}
  \begin{center}
    \includegraphics[width=0.495\textwidth]{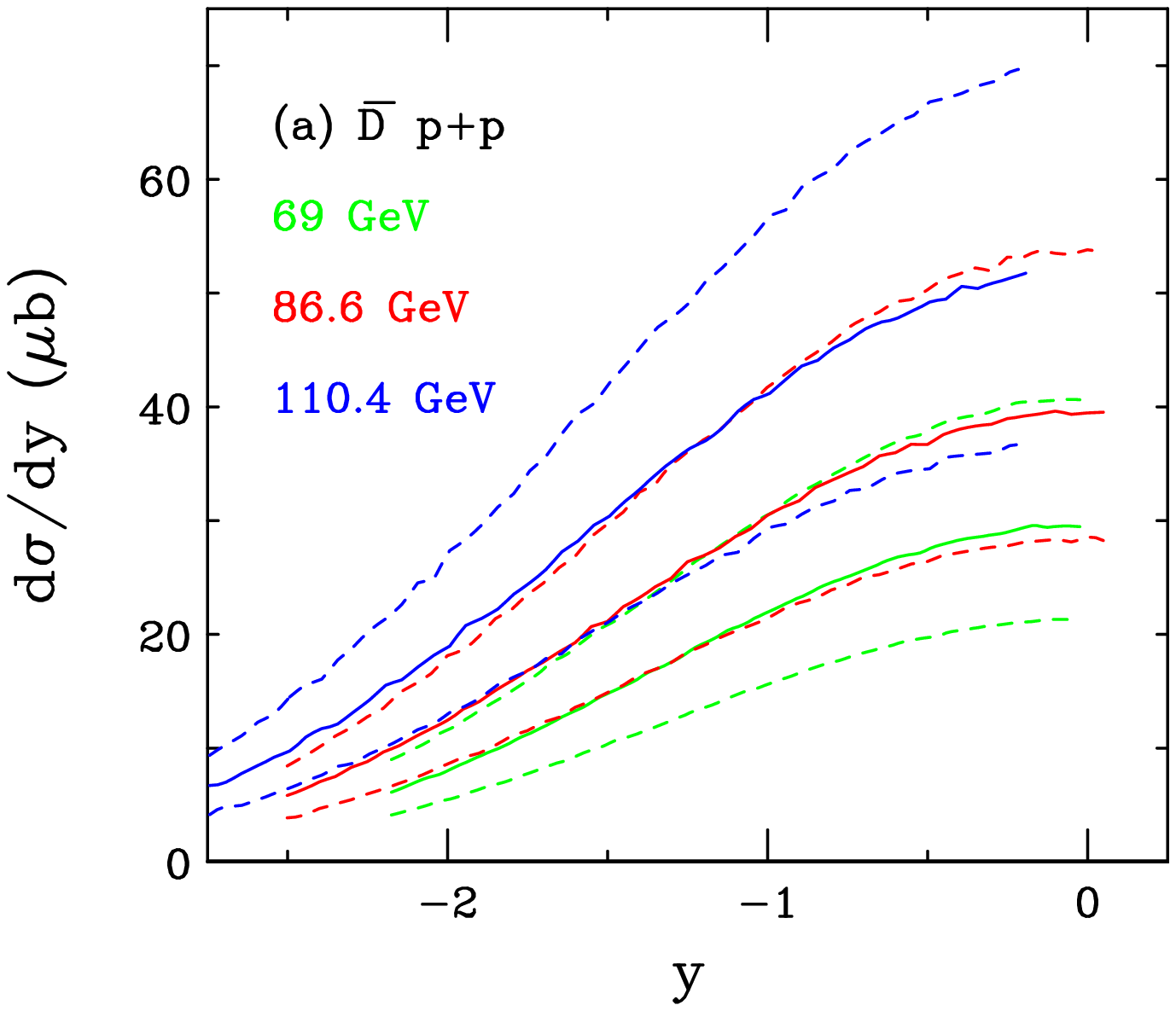}    
    \includegraphics[width=0.495\textwidth]{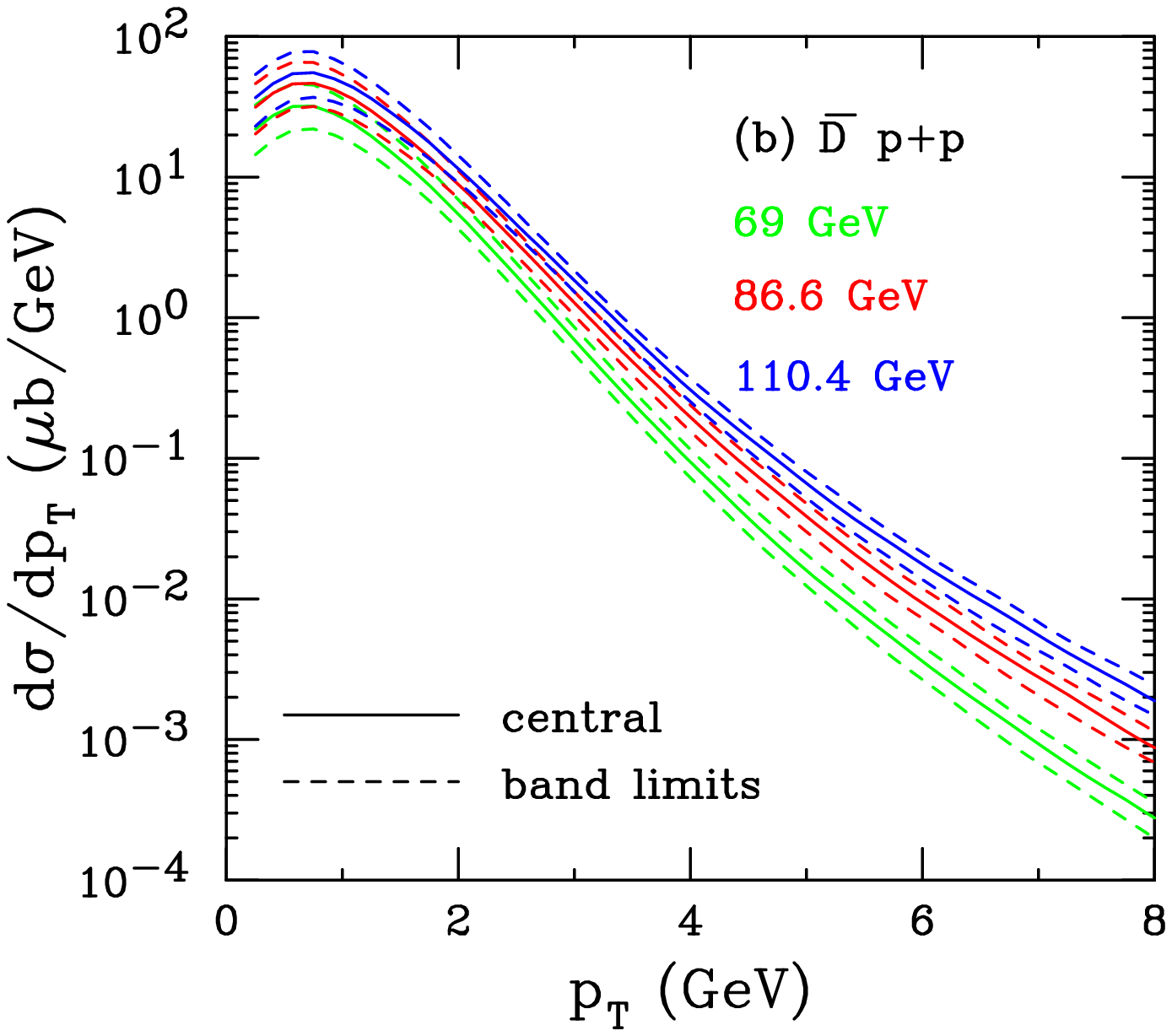}    
  \end{center}
  \caption[]{(Color online) The NLO $\overline D^0$ production cross sections 
    in $p+p$ collisions at $\sqrt{s} = 68.5$ (green), 86.6 (red), and 110.4~GeV
    (blue) as a function of rapidity (a) and $p_T$ (b), in the SMOG
    fixed-target acceptance, are shown.  The solid curves show
    the central values while the dashed curves outline the upper and lower
    limits of the uncertainty band.}
\label{d0_pp}
\end{figure}

Figure~\ref{d0_pp}
shows the rapidity and $p_T$ distributions for $\overline D^0$ production
in $p+p$ collisions at $\sqrt{s_{NN}} = 68.5$, 86.6, and 110.4~GeV.  The central value at each 
energy is given by the solid curve while the limits on the uncertainty bands 
are given by the dashed curves.  The rapidity distributions are shown in the SMOG range.  The $p_T$ distributions are integrated over these rapidity ranges.  

The uncertainty on the rapidity distribution, shown in Fig.~\ref{d0_pp}(a), at $\sqrt{s} = 68.5$~GeV is $\sim 33$\%, rising to $\sim 40$\% at 110.4~GeV.  The cross section clearly rises
with energy, such that the upper limit of one uncertainty band coincides with the central cross section at the next energy.  The increase in the cross section at low $p_T$ in Fig.~\ref{d0_pp}(b) is similar.  The main difference between the $p_T$ distributions at different energies is in the high $p_T$ tail which hardens with energy.  

\begin{figure}
  \begin{center}
    \includegraphics[width=0.495\textwidth]{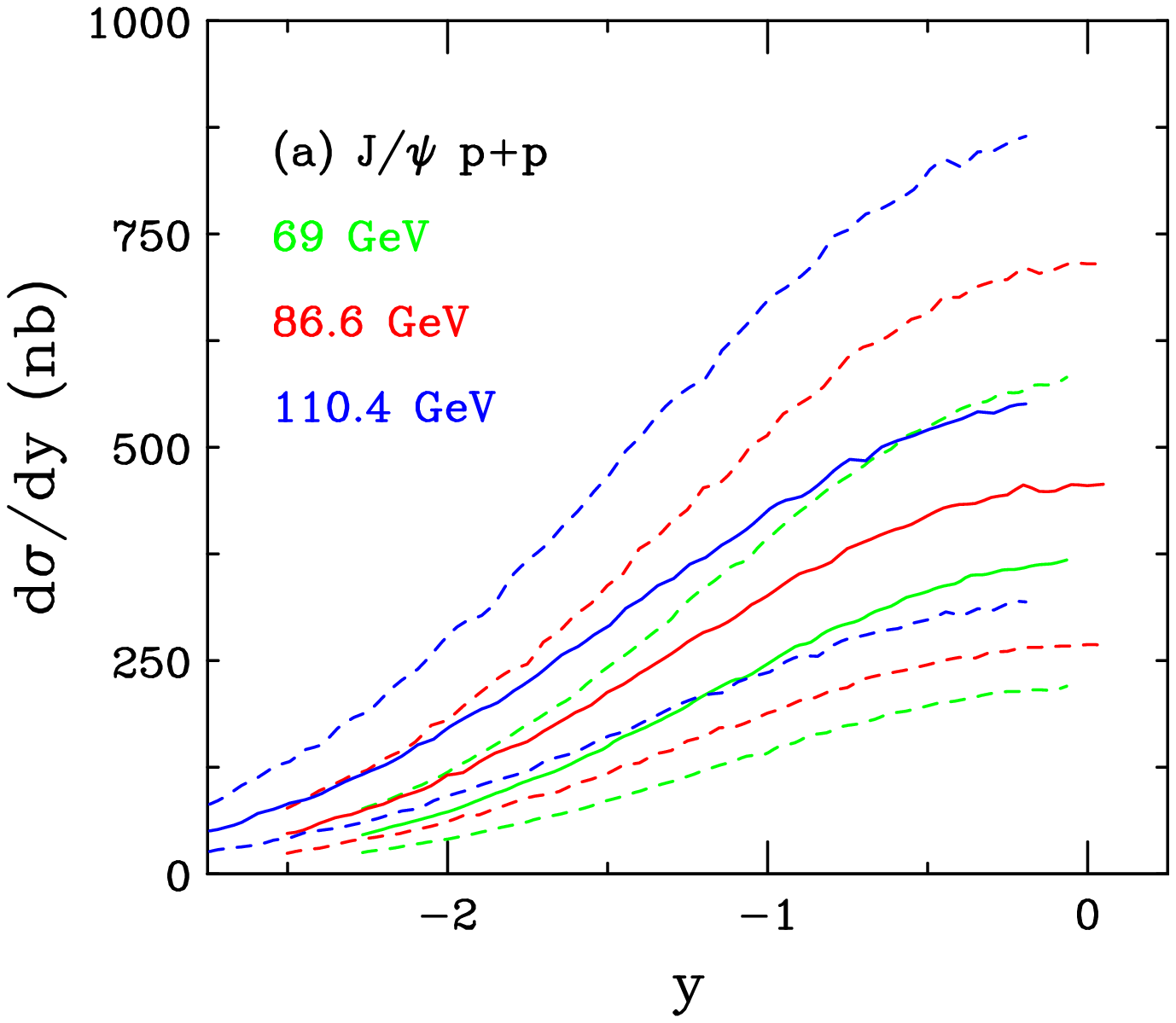}    
    \includegraphics[width=0.495\textwidth]{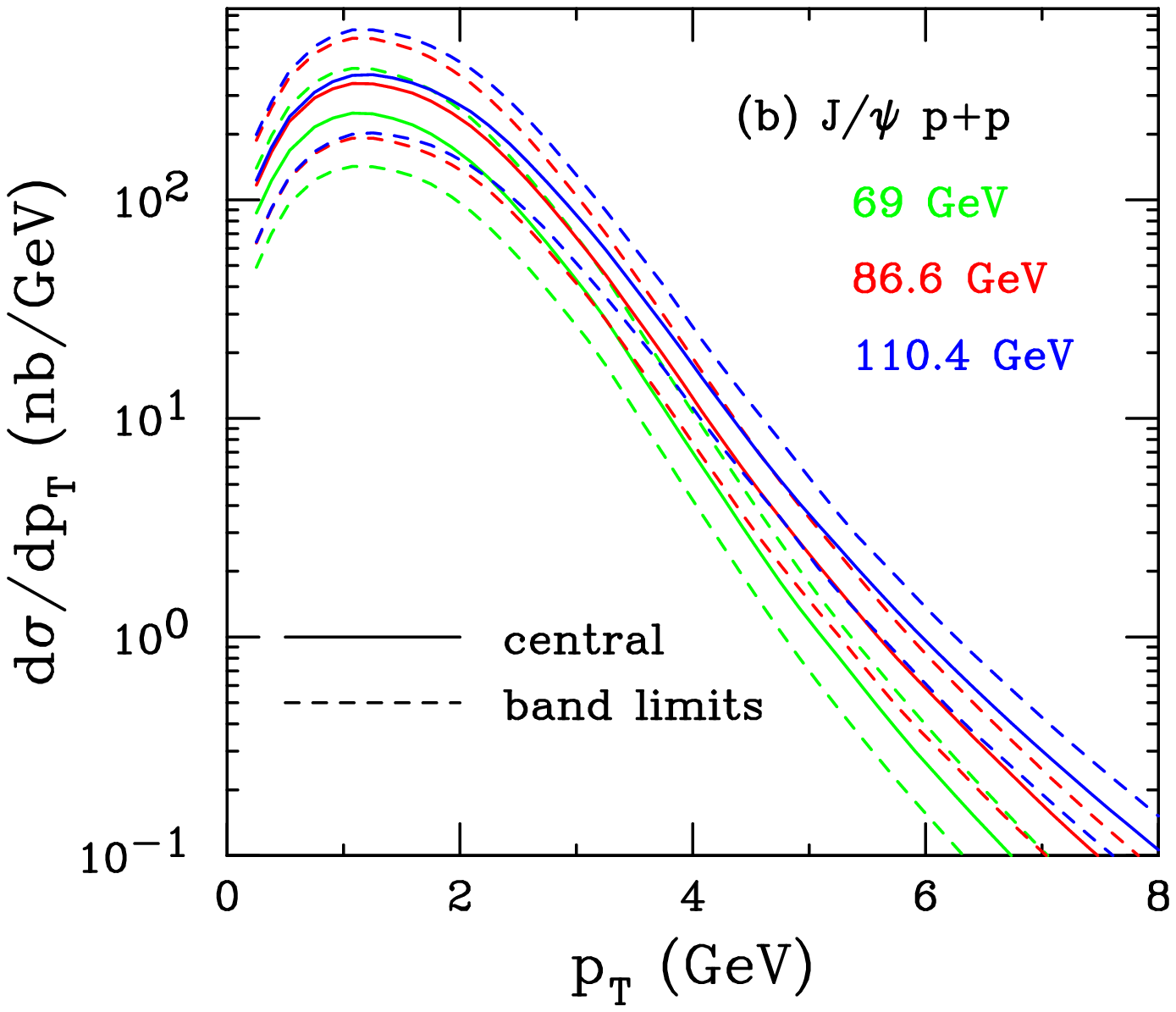}    
  \end{center}
\caption[]{The $J/\psi$ production cross sections in the CEM
  in $p+p$ collisions at $\sqrt{s} = 68.5$ (green), 86.6 (red), and 110.4~GeV
  (blue) as a function of rapidity (a) and $p_T$ (b), in the SMOG fixed-target
  acceptance, is shown.  The solid curves show
  the central values while the dashed curves outline the upper and lower limits
  of the uncertainty band.}
\label{cem_pp}
\end{figure}

Figure~\ref{cem_pp}
shows the rapidity and $p_T$ distributions for $J/\psi$ production
in $p+p$ collisions in the CEM at the same energies.  Similar trends are seen as for open charm production except the uncertainty at the most central rapidity, shown in Fig.~\ref{cem_pp}(a), is on the order of 60\%.  The $J/\psi$ $p_T$ distributions also harden with increased energy. It is notable, however, that they are harder overall than the $D$ meson distributions which fall off more steeply over the same $p_T$ range.

\section{Cold Nuclear Matter Effects}
\label{CNM}

Here nuclear modifications of the parton densities, transverse momentum broadening, 
and $J/\psi$ absorption in the nucleus, all cold nuclear matter effects in $p+A$
collisions are briefly discussed.  More details can be found in 
Ref.~\cite{RV_IC_EN}.  Only a brief discussion of each effect, specific to the SMOG 
setup, is included in this section.

The combined cold nuclear matter effects on perturbative QCD production of
open heavy flavor and $J/\psi$ are modified from Eqs.~(\ref{sigOHF}) and (\ref{sigCEM}) 
respectively, described in Sec.~\ref{pQCD}, are
\be
\sigma_{\rm OHF}(pA) & = & \sum_{i,j} 
\int_{4m^2}^\infty d\hat{s}
\int dx_1 \, dx_2~ F_i^p(x_1,\mu_F^2,k_T)~ F_j^A(x_2,\mu_F^2,k_T)~ 
\hat\sigma_{ij}(\hat{s},\mu_F^2, \mu_R^2) \, \, ,  \label{sigOHF_pA} \\
\sigma_{\rm CEM}(pA) & = & S_A^{\rm abs} F_C \sum_{i,j} 
\int_{4m^2}^{4m_H^2} d\hat{s}
\int dx_1 \, dx_2~ F_i^p(x_1,\mu_F^2,k_T)~ F_j^A(x_2,\mu_F^2,k_T)~ 
\hat\sigma_{ij}(\hat{s},\mu_F^2, \mu_R^2) \, \, , 
\label{sigCEM_pA}
\ee
where
\be
F_j^A(x_2,\mu_F^2,k_T) & = & R_j(x_2,\mu_F^2,A) f_j(x_2,\mu_F^2) G_A(k_T) \, \, \\
F_i^p(x_1,\mu_F^2,k_T) & = & f_i(x_1,\mu_F^2) G_p(k_T) \, \, .
\ee
The nuclear modifications of the parton distributions, $R_j(x_2,\mu_F^2,A)$ are 
discussed in Sec.~\ref{shad}.
The $k_T$ broadening in the nuclear target, $G_A(k_T)$, is discussed in Sec.~\ref{kTkick}.
Finally, $J/\psi$ absorption by nucleons, represented by the survival probability 
$S_A^{\rm abs}$, is described in Sec.~\ref{absorption}.

\subsection{Nuclear Effects on the Parton Densities}
\label{shad}

A number of global analyses have been made to
describe the modification as a function of $x$ and factorization scale $\mu_F$,
assuming collinear factorization and starting from a minimum scale, $\mu_F^0$.
Nuclear PDF (nPDF) effects generated in this scheme are
generally implemented by a parameterization as
a function of $x$, $\mu_F$ and $A$.  The $k_T$-independent
proton parton distribution
functions in Eqs.~(\ref{sigOHF}) and (\ref{sigCEM}) are replaced by the nPDFs,
\be
f_j^A(x_2,\mu_F^2) = R_j(x_2,\mu_F^2,A) f_j^p(x_2,\mu_F^2) \, \, ,
\ee

The NLO EPPS16 \cite{EPPS16} nPDF parameterization is employed in these
calculations for $A = 4$ (helium), 20 (neon) and 40 (argon).
EPPS16 has 20 fit parameters, for
41 total sets: one central set and 40 error sets.  The error sets are
determined by varying each parameter individually within one standard deviation
of its best fit value.
The uncertainties on $R_j(x_2,\mu_F^2,A)$ are calculated by summing the
excursions of each of the error sets from the central value in quadrature. 

The nPDF uncertainties on the $J/\psi$ and $D^0$ distributions are
obtained by calculating the perturbative cross sections at the central values assumed for
the charm mass and the factorization and renormalization scales employing the central EPPS16 set
as well as the 40 error sets and summing the differences in quadrature.  The resulting 
uncertainty bands deviate from the central cross section on the order of 20\%,
significantly less than the mass and scale uncertainties shown in
Figs.~\ref{d0_pp} and \ref{cem_pp}.

\begin{figure}
  \begin{center}
    \includegraphics[width=0.495\textwidth]{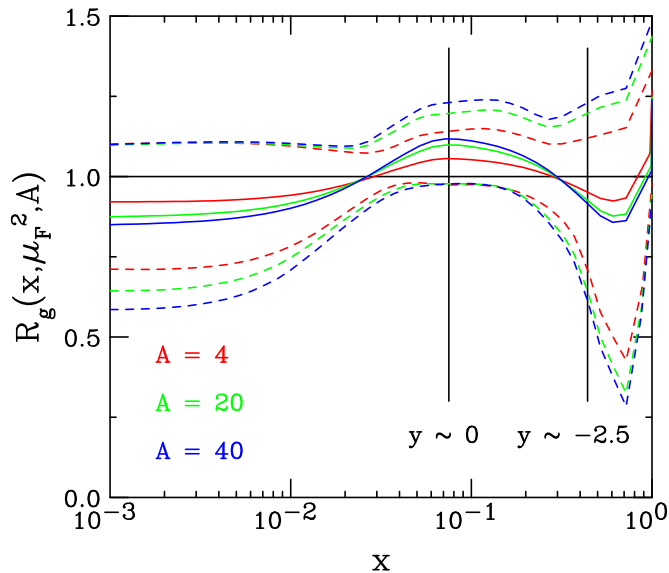}    
  \end{center}
  \caption[]{(Color online) The EPPS16 ratios, with uncertainties,
    are shown at the scale of the $J/\psi$ mass for gluons as a function of
    momentum fraction $x$. The central set is denoted by the solid curves
    while the dashed curves give the upper and lower limits of the uncertainty
    bands.  The results are given for $A = 4$ (red), 20 (green), and 40 (blue).
    The vertical lines indicate the $x$ range of the SMOG device,
    $0.075 < x < 0.44$.
  }
\label{shad_ratios}
\end{figure}

The EPPS16 ratios for gluons are
shown at the $J/\psi$ mass scale in Fig.~\ref{shad_ratios}.
The central sets, along with the uncertainty
bands, are shown for He ($A=4$), Ne ($A = 20$), and Ar ($A = 40$) targets.  
The EPPS16 gluon sets are shown as a function of $x$ for all three values of
$A$. The nPDF effects increase with $A$, as expected, but the effect is most significant for
lower values of $x$ than covered by the SMOG device.

The $x$ range covered by the SMOG device for near midrapidity to backward
rapidity, $y \sim -2.5$, is indicated by the vertical black bars.  The near
midrapidity value of $x$ is near the peak of the antishadowing region,
$x \sim 0.075$, while the most backward rapidity reaches into the EMC region,
$x \sim 0.44$.  The modification due to EPPS16 is smallest for $A = 4$ and
largest for $A = 40$, as expected.

The ratios are all shown for the $J/\psi$ mass,
$\mu_F = m_{J/\psi}$.  At this value of the factorization
scale, within a factor of three of
the minimum scale employed in the EPPS16 fits, $\mu_F = 1.3$~GeV, the
uncertainty band is relatively narrow in the $x$ region
spanned by SMOG.  The nPDF effect for the central set is a 10-15\% enhancement at $y \sim 0$
and a 5-10\% depletion at $y\sim -2.5$.  Including the range indicated by the error sets, 
increases the spread, particularly at negative rapidity, but the effect is still 
generally smaller than the mass and scale uncertainties on the $p+p$ cross sections themselves.

\subsection{$k_T$ Broadening}
\label{kTkick}

The effect and magnitude of intrinsic $k_T$ broadening on the $p_T$
distribution in $p+p$ collisions was discussed in Sec.~\ref{pQCD}.
Here further broadening due to the
presence of a nuclear target, single to multiple
parton scatterings in the nucleus, known as the Cronin effect \cite{Cronin} is described.  
The effect is implemented by replacing $g_p(k_T)$ by
$g_A(k_T)$ where instead of $\langle k_T \rangle_p$, $\langle k_T \rangle_A$ is employed
in Eq.~(\ref{intkt}).

The total broadening in a nucleus relative to a nucleon can be expressed as
\begin{eqnarray}
\langle k_T^2 \rangle_A = \langle k_T^2 \rangle_p +\delta k_T^2 \, \, ,
\end{eqnarray}
where $\delta k_T^2$ is \cite{HPC_pA,XNW_PRL}
\begin{eqnarray}
  \delta k_T^2 = (\langle \nu \rangle - 1) \Delta^2 (\mu) \, \, .
  \label{delkt2}  
\end{eqnarray}
The amount of broadening, $\Delta^2 (\mu)$, depends on the interaction
scale \cite{XNW_PRL} and the number of scatterings the incident proton
undergoes while passing through the nucleus \cite{RV_SeaQuest,RV_IC_EN},
\begin{eqnarray}
  \delta k_T^2 \approx (0.92 A^{1/3} - 1)\times 0.101 \, {\rm GeV}^2 \, \, .
  \end{eqnarray}
For helium, neon, and argon targets, $\delta k_T^2 = 0.05$, 0.15, and 0.22~GeV$^2$ 
respectively, giving an 
average broadening of
$\langle k_T^2 \rangle_A = 1.17$, 1.25, and 1.36~GeV$^2$ for the 
$p+{\rm He}$, $p+{\rm Ne}$ and $p+{\rm Ar}$ systems respectively. 

The effect of $k_T$ broadening in nuclei, relative to the proton, is to reduce
the ratio $p+A/p+p$ at low $p_T$ and enhance it at high
$p_T$.  The rapidity distributions are unaffected because they do not 
depend on $p_T$.  

\subsection{Nuclear Absorption of $J/\psi$ in $p+A$ Interactions}
\label{absorption}

In $p+A$ collisions, the proto-$J/\psi$ may interact with nucleons and be dissociated before 
it can escape the target, referred to as nuclear absorption.
The effect of nuclear absorption on the $J/\psi$ 
production cross section in
$p+A$ collisions may be expressed as \cite{rvrev}
\begin{eqnarray}
  \sigma_{pA} = \sigma_{pN} S_A^{\rm abs} & = & \sigma_{pN} \int d^2b \,
  \int_{-\infty}^{\infty}\, dz \, \rho_A (b,z) S^{\rm abs}(b) \\
& = & \sigma_{pN} \int d^2b \,  \int_{-\infty}^{\infty}\, dz \, \rho_A (b,z)
 \exp \left\{
-\int_z^{\infty} dz^{\prime} \rho_A (b,z^{\prime}) \sigma_{\rm abs}(z^\prime
-z)\right\} \, \, , 
\label{sigfull}
\end{eqnarray} where $b$ is the impact parameter, $z$ is the $c \overline c$
production point, $S^{\rm abs}(b)$ is the nuclear absorption survival 
probability, and $\sigma_{\rm abs}(z^\prime -z)$ 
is the nucleon absorption cross section. 
The absorption cross section here is assumed to be constant at a given energy. 
It is written as a function of the
path length through the nucleus in Eq.~(\ref{sigfull}) because other functional
forms may be chosen, see {\it e.g.}\ Ref.~\cite{RV_HeraB}.

The energy dependence of $\sigma_{\rm abs}$
was studied in  Ref.~\cite{LWV}.
Using those results and following Ref.~\cite{RV_IC_EN}, values of 
$\sigma_{\rm abs} = 4$, 3.5, and 3~mb are used  
at $\sqrt{s_{NN}} = 68.5$, 86.6 and 110.4~GeV.

Absorption by comoving particles, whether characterized
as hadrons or partons \cite{Elena}
has not been included.  It has the same nuclear dependence in minimum
bias collisions as absorption by nucleons \cite{SGRV1990}.

\section{Intrinsic Charm}
\label{IC}

As first proposed in 1980, the proton wave function in QCD can be represented as a
superposition of Fock state fluctuations, {\it e.g.}\ $\vert uudg
\rangle$, $\vert uud q \overline q \rangle$, $\vert uud Q \overline Q \rangle$,
\ldots of the $\vert uud \rangle$ state.  When charm quarks, $Q = c$, are part of the state, it is referred to as intrinsic charm, or IC.
If a proton in such a state scatters in a target, the
coherence of the Fock components is broken and the fluctuations can
hadronize \cite{intc1,intc2,BHMT}.  These
intrinsic $Q
\overline Q$ Fock states are dominated by configurations with
equal rapidity constituents, so that the heavy quarks carry a large
fraction of the proton momentum \cite{intc1,intc2}.  (While proton
projectiles are emphasized here, any hadron wave function can be so described.)
 
The frame-independent probability distribution of a $5$-particle IC
Fock state in the proton is 
\be
dP_{{\rm ic}\, 5} = P_{{\rm ic}\,5}^0
N_5 \int dx_1 \cdots dx_5 \int dk_{x\, 1} \cdots dk_{x \, 5}
\int dk_{y\, 1} \cdots dk_{y \, 5} 
\frac{\delta(1-\sum_{i=1}^5 x_i)\delta(\sum_{i=1}^5 k_{x \, i}) \delta(\sum_{i=1}^5 k_{y \, i})}{(m_p^2 - \sum_{i=1}^5 (\widehat{m}_i^2/x_i) )^2} \, \, ,
\label{icdenom}
\ee
where $i = 1$, 2, 3 are the light quarks ($u$, $u$, $d$)
and $i = 4$ and 5 are the $c$ and $\overline c$ quarks respectively.
The factor $N_5$ normalizes the
$|uud c \overline c \rangle$ probability to unity and $P_{{\rm ic}\, 5}^0$
scales the unit-normalized
probability to the assumed probability of IC in the proton.
The delta functions in Eq.~(\ref{icdenom}) conserve longitudinal ($z$) and
transverse ($x$ and $y$) momentum.  The denominator of Eq.~(\ref{icdenom}) is
minimized when the charm quarks carry the largest fraction of the
proton longitudinal momentum, $\langle x_c \rangle > \langle x_q \rangle$.
The default values of the quark masses and $k_T$-integration ranges in
Eq.~(\ref{icdenom}) are $m_c = 1.27$~GeV, $m_q = 0.3$~GeV,
$k_q^{\rm max} = 0.2$~GeV and $k_c^{\rm max} = 1.0$~GeV.  Changing the $k_T$ integration 
range of the partons in the state does
not strongly affect the $y$ or $p_T$ distributions of the produced mesons.
(Note that the constitutent quark masses are used for the light quarks.)

Additional delta functions can be employed to describe hadronization by simple
coalescence when the Fock state is disrupted.  For example, the $J/\psi$ $x_F$
distribution can be calculated
by the addition of the delta functions,
$\delta(x_F - x_c - x_{\overline c})$, in the longitudinal, $z$, direction.
The summed $x_c$ and $x_{\overline c}$ momentum fractions are equivalent to the
$x_F$ of the $J/\psi$ assuming that it is brought on-shell by a soft scattering
with the target.  Similarly, the $J/\psi$ 
$p_T$ distribution is described by $\delta(p_T - k_{x \, c} - k_{x \, \overline c}) \delta(k_{y \, c} + k_{y\, \overline c})$ where the $J/\psi$ $p_T$ is chosen to be
along the $x$ direction for simplicity and without loss of generality.  

Likewise $\overline D$ mesons ($D^-(\overline c d)$ and
$\overline D^0 (\overline c u)$) mesons can be directly produced from the disrupted Fock state
employing $\delta(x_F - x_{\overline c} - x_i)$ for the $\overline D$ $x_F$ and
$\delta(p_T - k_{x \, \overline c} - k_{x \, i}) \delta(k_{y \, i} + k_{y\, \overline c})$ for the $p_T$
where the light parton $i$ can be either a $u$ or $d$ quark.  
The remaining
partons in the state could coalesce into a $\Lambda_c(udc)$ with a
$\overline D^0$ or a $\Sigma_c^{++}(uuc)$ with a $D^-$.  Thus the 5-particle proton IC state could
produce $\overline D^9 \Lambda_c$ or $D^- \Sigma_c^{++}$ through coalescence.

The $J/\psi$ and $\overline D$ $x_F$ and $p_T$ distributions,
integrated over all phase space, are independent of the proton energy.  The
$p_T$ distribution from IC only varies when phase space cuts are
considered, as shown in Refs.~\cite{RV_SeaQuest,RV_IC_EN}.  The
rapidity distribution, however, depends strongly on $\sqrt{s}$ because $x_F = (2m_T/\sqrt{s}) \sinh y$.  Thus even
though the $x_F$ distribution is invariant with $\sqrt{s}$, $m_i$
and $k_T$-integration range, the rapidity distribution is not \cite{RV_IC_EN}.
Indeed, the $x_F$
distribution depends only weakly on the heavy quark mass~\cite{RVSJB_psipsi,ANDY}.

The $J/\psi$ and $\overline D$ $p_T$ distributions have long tails at large $p_T$ rather
than decreasing strongly with increasing $p_T$ as the perturbative QCD calculations do
\cite{RV_IC_EN}.  This is due to the nature of Eq.~(\ref{icdenom}).  The requirement that
the heavy quarks have higher velocities not only gives results in more forward production
at low $p_T$ but also allows them to carry most of the transverse momentum at low $x$.
This also explains why the $p_T$ distributions are suppressed at low $p_T$ when only
the midrapidity (generally low $x$) region is considered \cite{RV_IC_EN}.
There is thus more than one way to satisfy minimization of the denominator of 
Eq.~(\ref{icdenom}) while still satisfying momentum conservation.

In the case of open charm, the preference for
$\overline D$ production in a 5-particle IC state makes the $\overline D$ a ``leading''
particle relative to $D^+ (c \overline d)$ and $D^0 (c \overline u)$.  These ``non-leading"
$D$ mesons could only be produced from a 5-particle IC state by standard fragmentation and would thus be produced at lower
$x_F$ or rapidity than the $\overline D$ mesons.  In order to produce a $D$ meson by 
coalescence from a IC state, as is the case for $\overline D$ in the $|uud c \overline c \rangle$
state, a higher particle-number Fock state is required, such as the seven-particle state
$|uud c \overline c d \overline d\rangle$, resulting in $D^+$ production \cite{tomg}.  

Previous studies of leading $D$ meson production, including asymmetries
between $D^+$ and $D^-$ production in fixed-target $\pi^- A$ interactions have
shown significant differences between leading and non-leading production
\cite{E769,E791,WA82}.  These asymmetries have been reproduced by IC
\cite{RVSJB_asymm} as well as by string-breaking mechanisms such as in
PYTHIA \cite{Norrbin}.  Later work has
shown that there can be an asymmetry in $c$ and $\overline c$ distributions
themselves \cite{Sufian:2020coz}.  This asymmetry arises from QCD diagrams where
two gluons from  two different valence quarks in the nucleon couple to a heavy
quark pair, $gg \rightarrow Q\overline Q$, with charge conjugation $C=+1$
\cite{Stan_review}.  This amplitude interferes with QCD diagrams where an
odd number of gluons attach to the heavy quark pair, {\it e.g.}
$ g\rightarrow Q\overline Q$ and $ggg\rightarrow Q \overline Q$, with $C=-1$.
The interference of amplitudes with the same final state but different $C$ for
the $Q \overline Q$ pair produces asymmetric distribution functions.  An
analogous interference term is seen in $e^-$ and $e^+$ distributions in
$e^+ e^-$ pair production \cite{Brodsky:1968rd}.  
In this work, it is assumed that the asymmetry between
leading and non-leading $\overline D$ and $D$ meson arises from 
their manifestation as final-state charm hadrons from the
5-particle IC state considered here.  The asymmetry due to charge conjugation
effects is not taken into account.

Note that if higher Fock states are considered, only 
equal rapidity $D$ and $\overline D$ mesons would be produced from
these states and at a lower average
momentum fraction for both the $D$ and $\overline D$.  The probability to
produces these higher Fock states would also be reduced,
see {\it e.g.}\ Ref.~\cite{tomg}.  
Here only the 5-particle proton Fock state is
considered since it gives the most forward $J/\psi$
and $\overline D$ production from IC.  This assumption also maximizes the
possible asymmetry between $D$ and $\overline D$ production from IC.

The IC production cross section from the
$|uudc \overline c \rangle$ state can be written as 
\be
\sigma_{\rm ic}(pp) = P_{{\rm ic}\, 5} \sigma_{p N}^{\rm in}
\frac{\mu^2}{4 \widehat{m}_c^2} \, \, .
\label{icsign}
\ee
The factor of $\mu^2/4 \widehat{m}_c^2$ is from the soft
interaction which breaks the coherence of the Fock state.  Here 
$\mu^2 = 0.1$~GeV$^2$ is assumed, see Ref.~\cite{VBH1}, and an inelastic
$pN$ cross section, $\sigma_{pN}^{\rm in} = 30$~mb, is employed.  Although
$\sigma_{pN}^{\rm in}$ can change slowly with $\sqrt{s}$, it is held constant here.

Equation~(\ref{icsign}) is used for open charm production,
$\sigma_{\rm ic}^{\overline D}(pp) = \sigma_{\rm ic}(pp)$. The
$J/\psi$ cross section from the same IC state
is calculated by scaling Eq.~(\ref{icsign}) by the 
factor $F_C$ used in the CEM calculation in
Eq.~(\ref{sigCEM}),
\be
\sigma_{\rm ic}^{J/\psi}(pp) = F_C \sigma_{\rm ic}(pp) \, \, .
\label{icsigJpsi}
\ee

The nuclear, $A$, dependence of the IC cross section is assumed to be that extracted for the nuclear surface-like component of $J/\psi$
production by the NA3 Collaboration \cite{NA3}.  The $A$
dependence is the same for both open charm and $J/\psi$,
\be 
\sigma_{\rm ic}^{\overline D}(pA) & = & \sigma_{\rm ic}^{\overline D}(pp) \, A^\beta \, \, , \label{icsigD_pA} \\
\sigma_{\rm ic}^{J/\psi}(pA) & = & \sigma_{\rm ic}^{J/\psi}(pp) \, A^\beta \, \, , 
\label{icsigJpsi_pA}
\ee
with $\beta = 0.71$ \cite{NA3}.

Several values of $P_{{\rm ic}\, 5}^0$ have been employed previously. 
Here, a value of 1\% is assumed to maximize the potential effect.
The form of IC
postulated by Brodsky and collaborators in Refs.~\cite{intc1,intc2}, used in 
Eq.~(\ref{icdenom}), has been
adopted.  Other variants of the IC distribution in the proton
have been proposed, including meson-cloud models where the proton fluctuates into a
$\overline D(u \overline c) \Lambda_c (udc)$ 
\cite{Paiva:1996dd,Neubert:1993mb,Steffens:1999hx,Hobbs:2013bia} and a
sea-like distribution \cite{Pumplin:2007wg,Nadolsky:2008zw}.  The $\overline D$ distribution here is
similar to that in the meson-cloud model while a sea-like distribution would not produce leading $D$ mesons and, indeed, would not result in forward charm production.

IC has been included in global analyses of the proton parton densities
\cite{Pumplin:2007wg,Nadolsky:2008zw,Dulat:2013hea,Jimenez-Delgado:2014zga,NNPDF_IC}.
The range of $P_{{\rm ic}\, 5}^0$
explored here is consistent with the upper limits of the results of these
analyses.  For more details of these other works, see the review of
Ref.~\cite{IC_rev}.  Since the discussion of Ref.~\cite{IC_rev}, new work has appeared claiming
evidence of IC at the 1\% level \cite{LHCb_intc,NNPDF}, as mentioned in the introduction. This finding remains contentious \cite{Guzzi:2022rca}.
(See Ref.~\cite{Blumlein} for a discussion of a possible
kinematic constraint on intrinsic charm in deep-inelastic scattering.)
New evidence for a finite charm quark asymmetry in the
nucleon wavefunction from lattice gauge theory, consistent with IC, was presented in Ref.~\cite{Sufian:2020coz}.  See also the recent
review in Ref.~\cite{Stan_review} for
more applications of intrinsic heavy quark states.

\section{Results}
\label{comp}

In this section, the rapidity and $p_T$ distributions for $J/\psi$ and open charm production in $p+A$ collisions are compared to those in $p+p$ collisions.  The cold nuclear matter effects introduced in Sec.~\ref{CNM} are included, as well as the IC contribution described in Sec.~\ref{IC}.  The calculations are compared to the SMOG data.  The asymmetry between $\overline D^0$ and $D^0$ mesons is also computed.

The cross sections for $\overline D$ and $J/\psi$ production in $p+p$ collisions
including the perturbative QCD and IC contributions are:
\be
\sigma_{pp}^{\overline D} & = & \sigma_{\rm OHF}(pp) + \sigma_{\rm ic}^{\overline D}(pp)
\label{sig_pp_Dsum} \\
\sigma_{pp}^{J/\psi} & = & \sigma_{\rm CEM}(pp) + \sigma_{\rm ic}^{J/\psi}(pp) \, \, .
\label{sig_pp_Jsum} 
\ee
Here $\sigma_{\rm OHF}(pp)$ and $\sigma_{\rm CEM}(pp)$ are defined in Eqs.~(\ref{sigOHF}) and (\ref{sigCEM}) respectively.
The IC contributions can be found in Eq.~(\ref{icsign}) for
$\overline D$ while
$\sigma_{\rm ic}^{J/\psi}(pp)$ is given in Eq.~(\ref{icsigJpsi}).

Likewise, the cross sections for $\overline D$ and $J/\psi$ production in $p+A$ collisions are:
\be
\sigma_{pA}^{\overline D} & = & \sigma_{\rm OHF}(pA) + \sigma_{\rm ic}^{\overline D}(pA)
\label{sig_pA_Dsum} \\
\sigma_{pA}^{J/\psi} & = & \sigma_{\rm CEM}(pA) + \sigma_{\rm ic}^{J/\psi}(pA) \, \, .
\label{sig_pA_Jsum} 
\ee
Now $\sigma_{\rm OHF}(pA)$ and $\sigma_{\rm CEM}(pA)$ are defined in Eqs.~(\ref{sigOHF_pA}) and (\ref{sigCEM_pA}) respectively while the $\overline D$ and $J/\psi$ IC cross sections are in Eqs.~(\ref{icsigD_pA}) and (\ref{icsigJpsi_pA}).

\begin{figure}
  \begin{center}
    \includegraphics[width=0.4\textwidth]{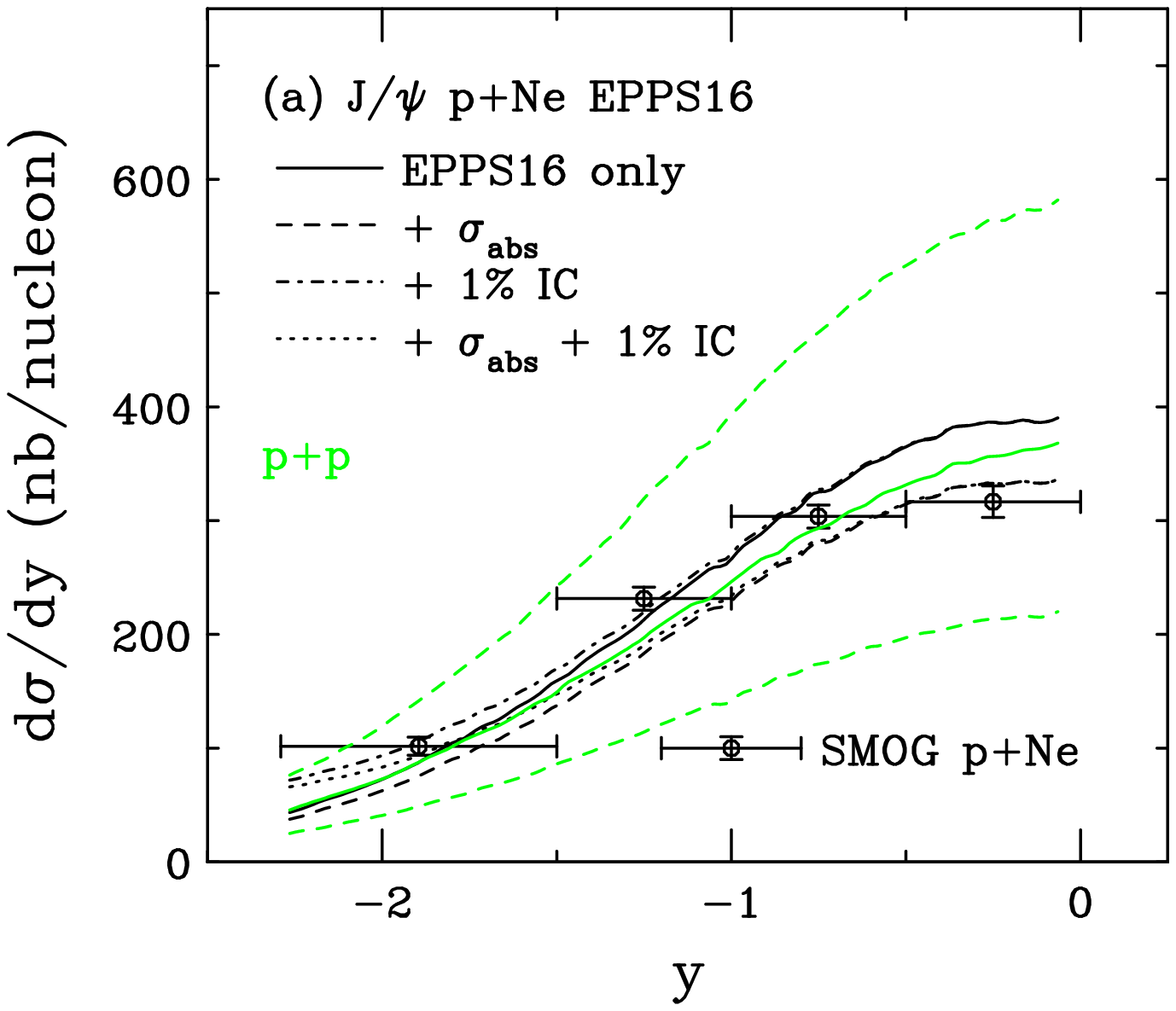}
    \includegraphics[width=0.4\textwidth]{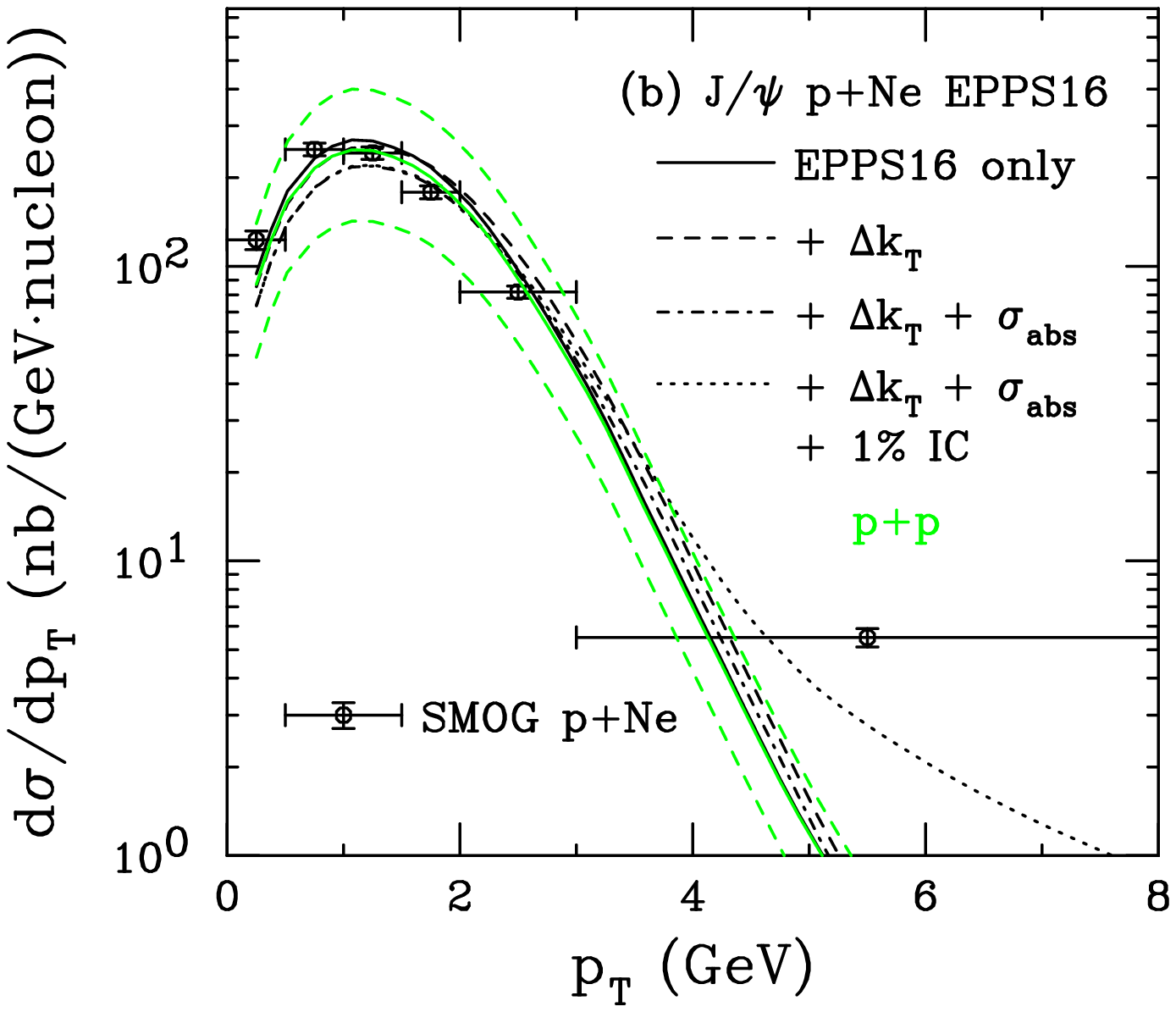} \\
    \includegraphics[width=0.4\textwidth]{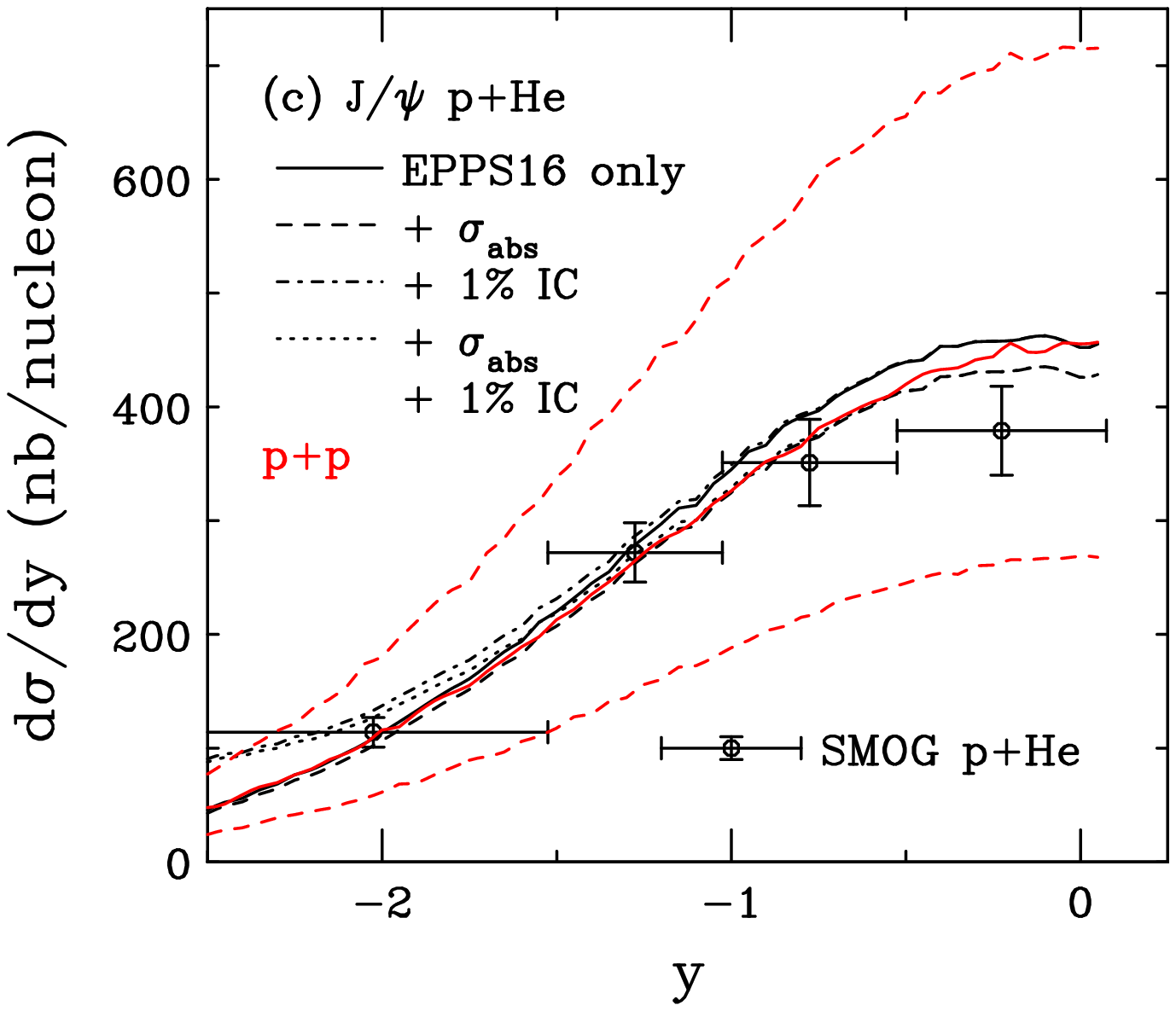}
    \includegraphics[width=0.4\textwidth]{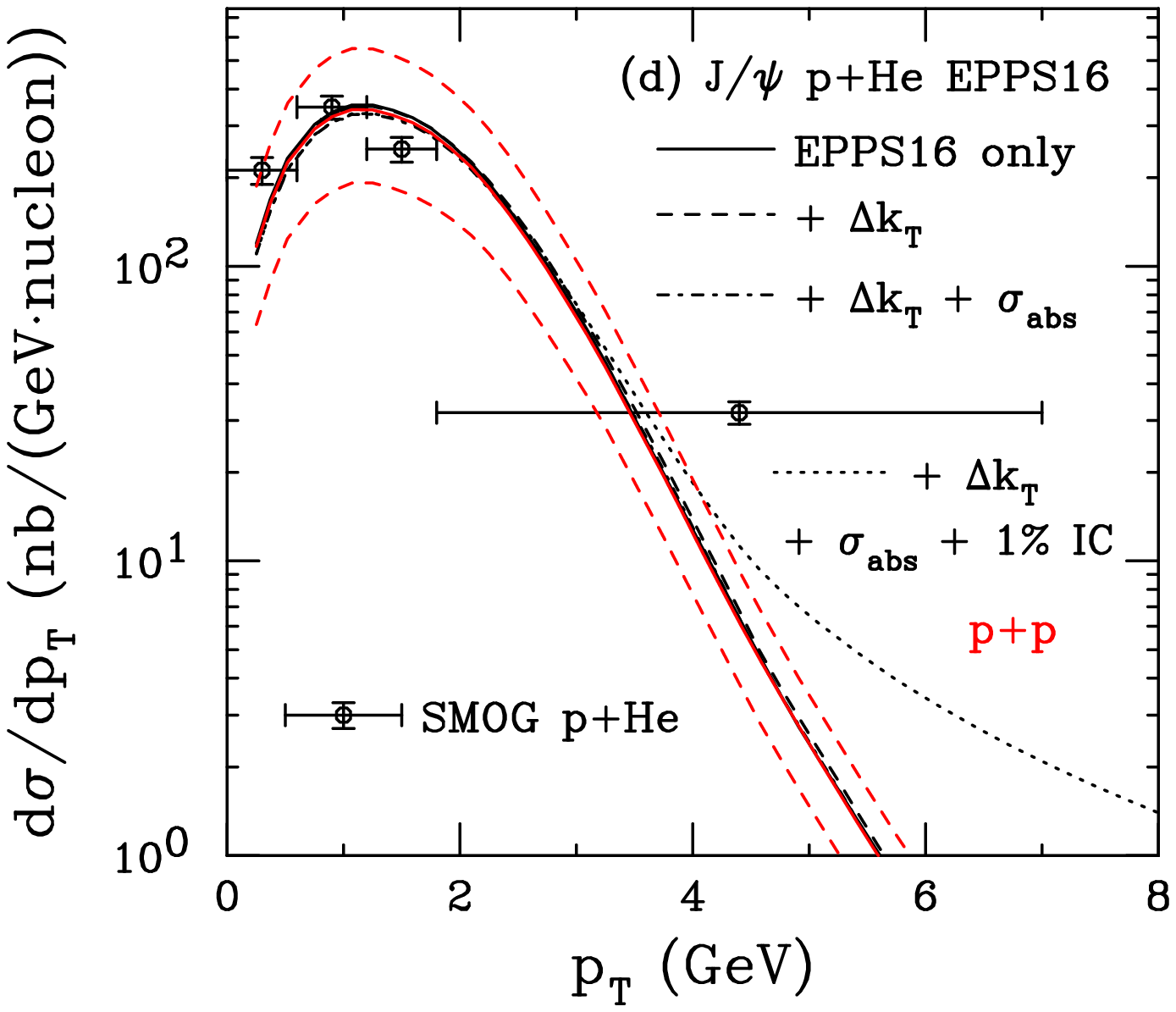} \\
    \includegraphics[width=0.4\textwidth]{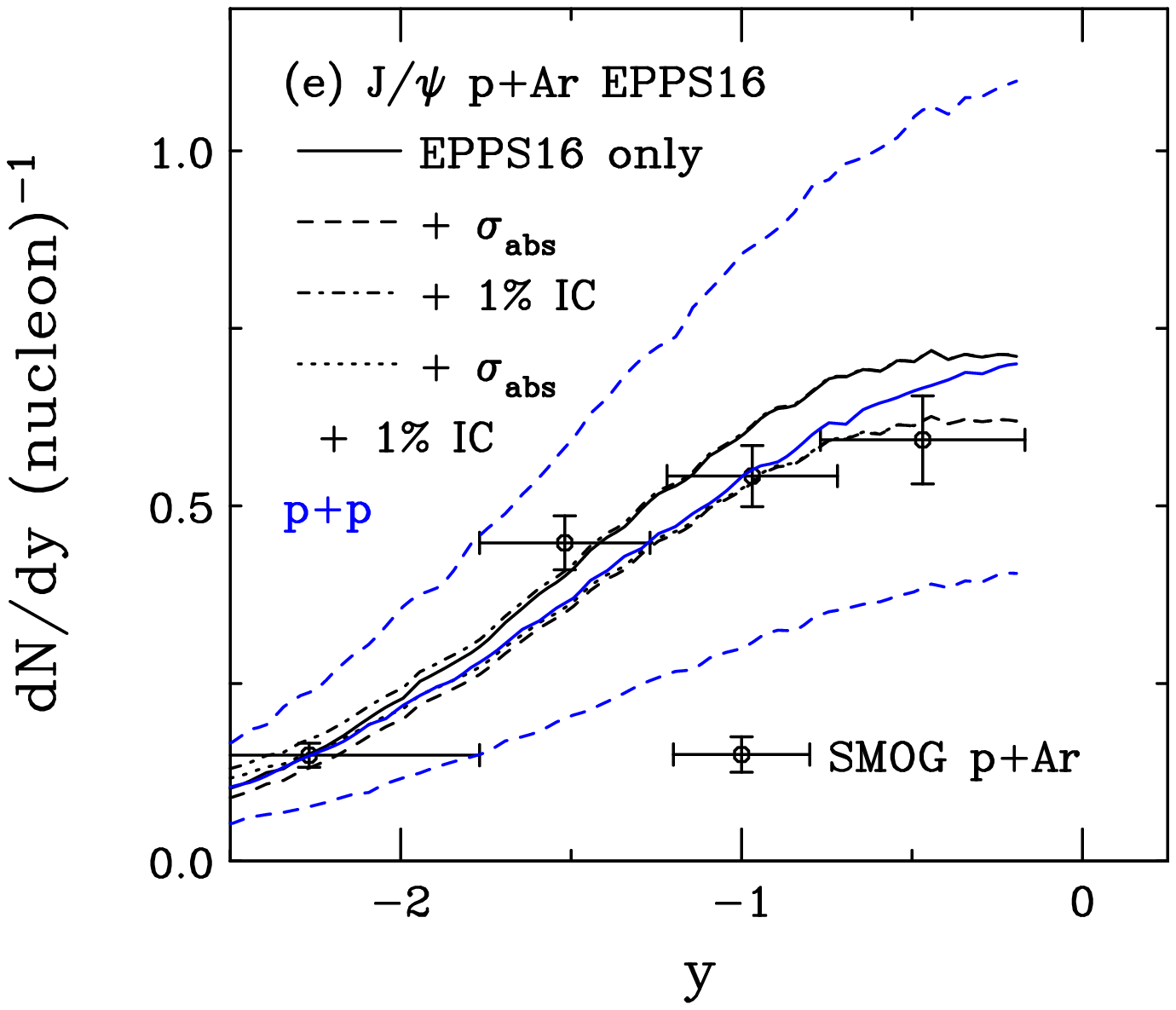}
    \includegraphics[width=0.4\textwidth]{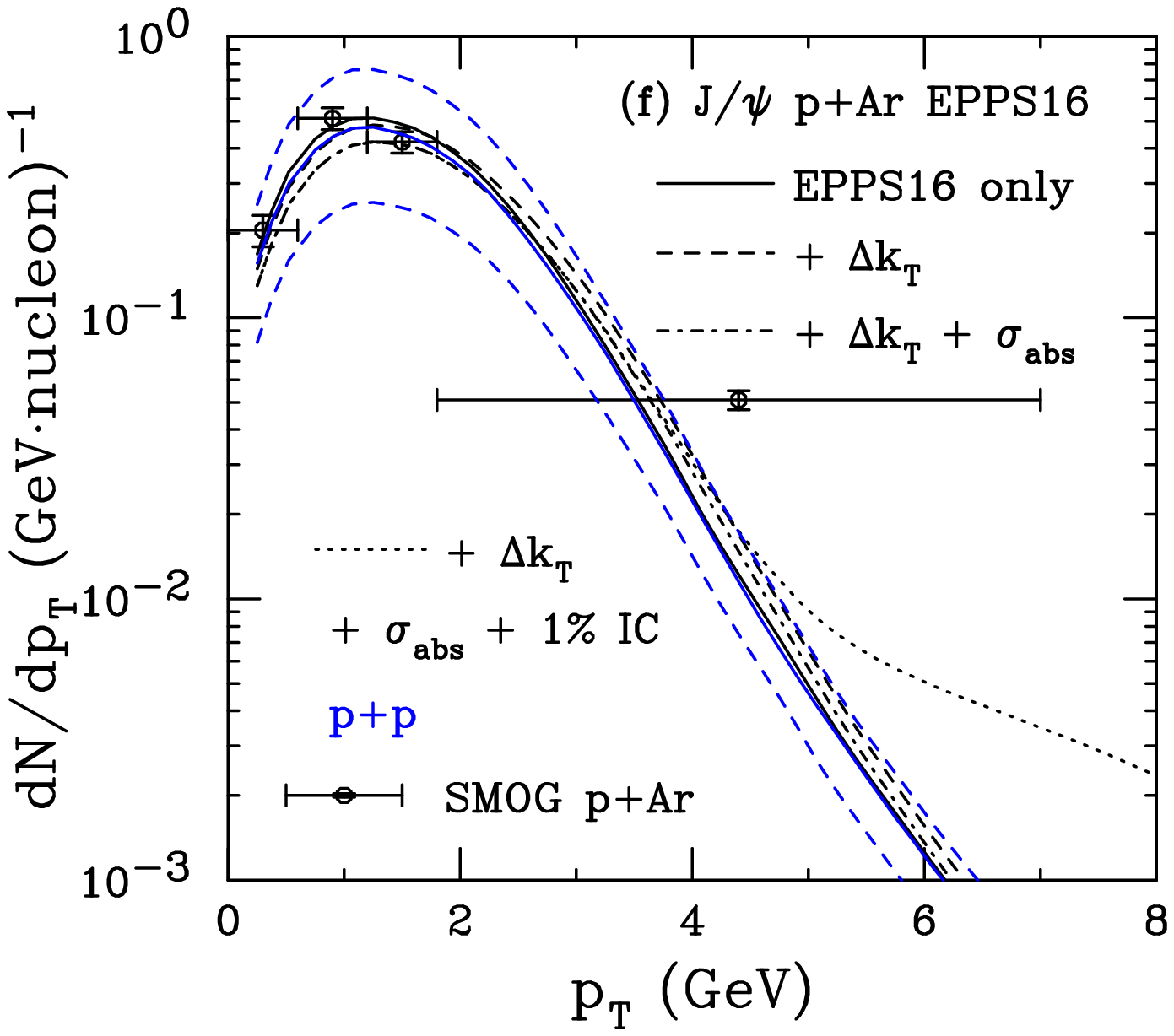}
  \end{center}
  \caption[]{ The  $J/\psi$ cross section as a function
    of rapidity in (a), (c), (e) and transverse momentum in (b), (d), (f)
    for $p+{\rm Ne}$ collisions at $\sqrt{s_{NN}} = 68.5$~GeV in (a) and (b);
    $p+{\rm He}$ collisions at $\sqrt{s_{NN}} = 86.6$~GeV in (c) and (d); and
    $p+{\rm Ar}$ collisions at $\sqrt{s_{NN}} = 110.4$~GeV in (e) and (f).
    The black curves are the $p+A$ calculations.  The colored curves (solid and
    dashed) show the CEM $p+p$ calculations (no IC) at the same energy for the central
    value and the limits of the uncertainty band.  
    The $p+A$ rapidity distributions are shown for EPPS16 only (solid); EPPS16
    with absorption (dashed); EPPS16 and $P_{\rm ic \, 5}^0 = 1$\% (dot-dashed);
    and EPPS16, absorption, and $P_{\rm ic \, 5}^0 = 1$\% (dotted).  The $p_T$
    distributions show EPPS16 only (solid); EPPS16 with $k_T$ kick (dashed);
    EPPS16, absorption, and $k_T$ kick (dot-dashed); and EPPS16, absorption,
    $k_T$ kick and $P_{\rm ic \, 5}^0 = 1$\% (dotted).  The SMOG $p+{\rm Ne}$
    data are from Ref.~\cite{SMOGpNeJ} while the $p+{\rm He}$ and $p+{\rm Ar}$
    data are from Ref.~\cite{SMOG}.
    }
\label{fig:JpAdists}
\end{figure}

\begin{figure}
  \begin{center}
    \includegraphics[width=0.4\textwidth]{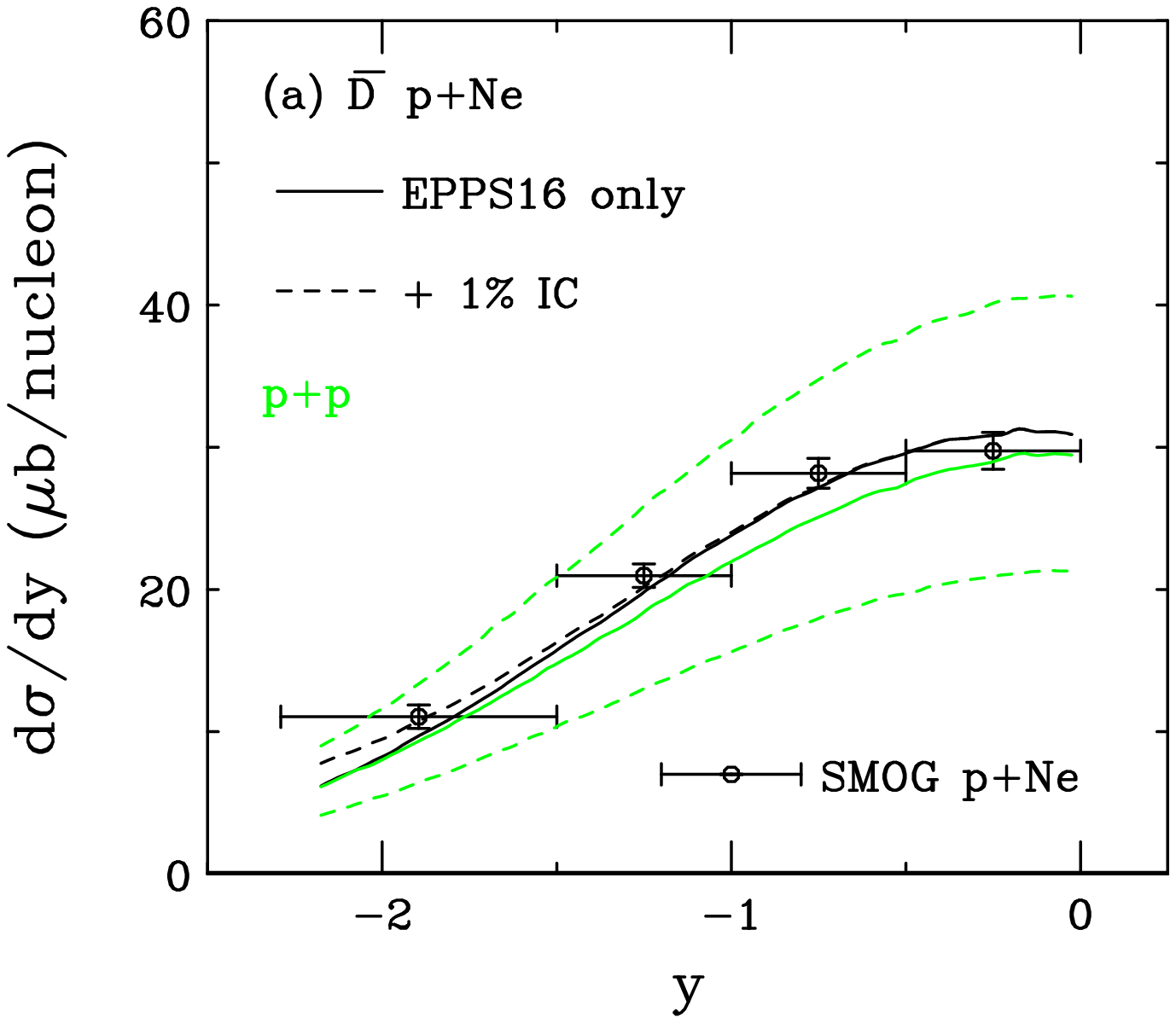}
    \includegraphics[width=0.4\textwidth]{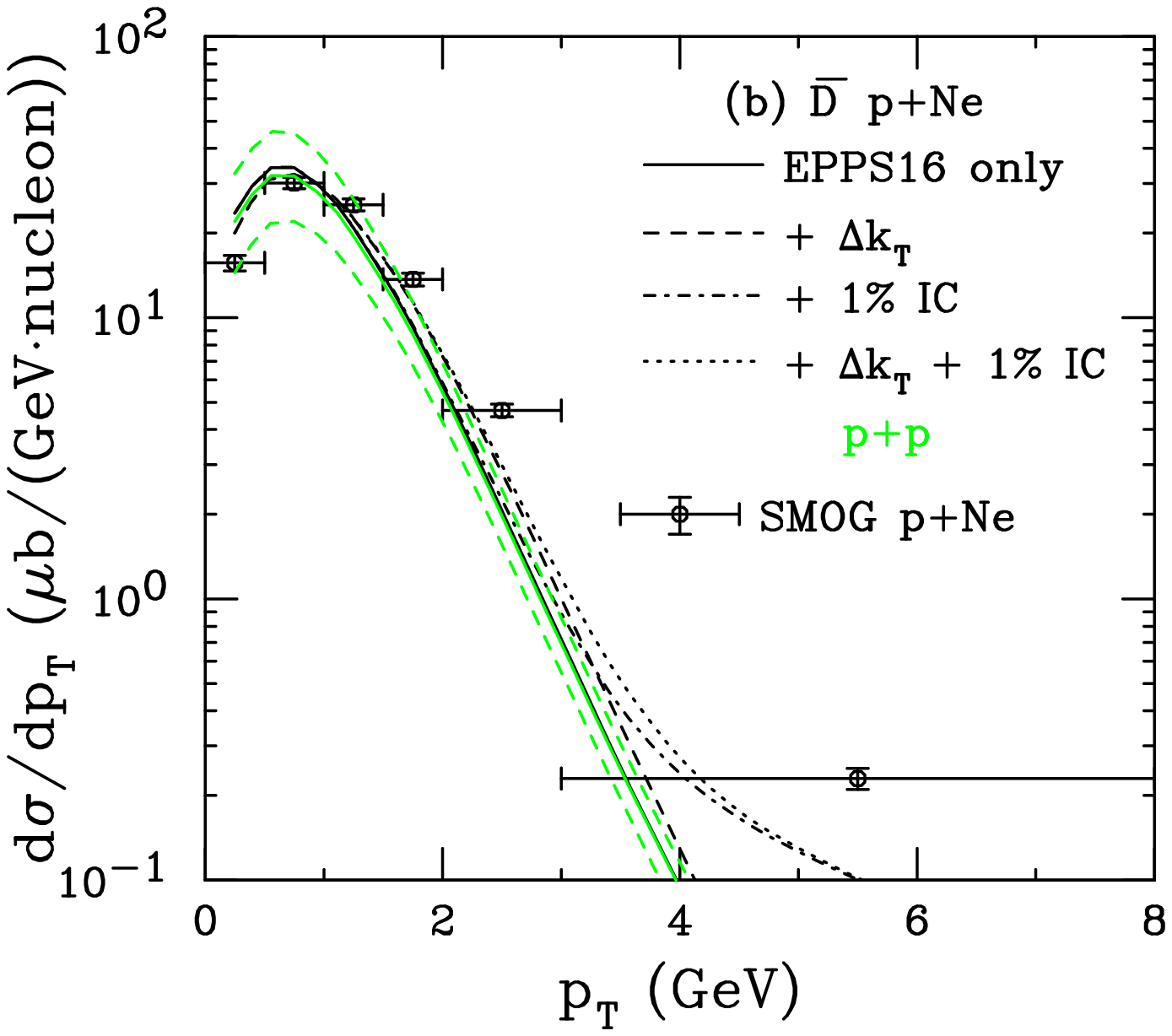} \\
    \includegraphics[width=0.4\textwidth]{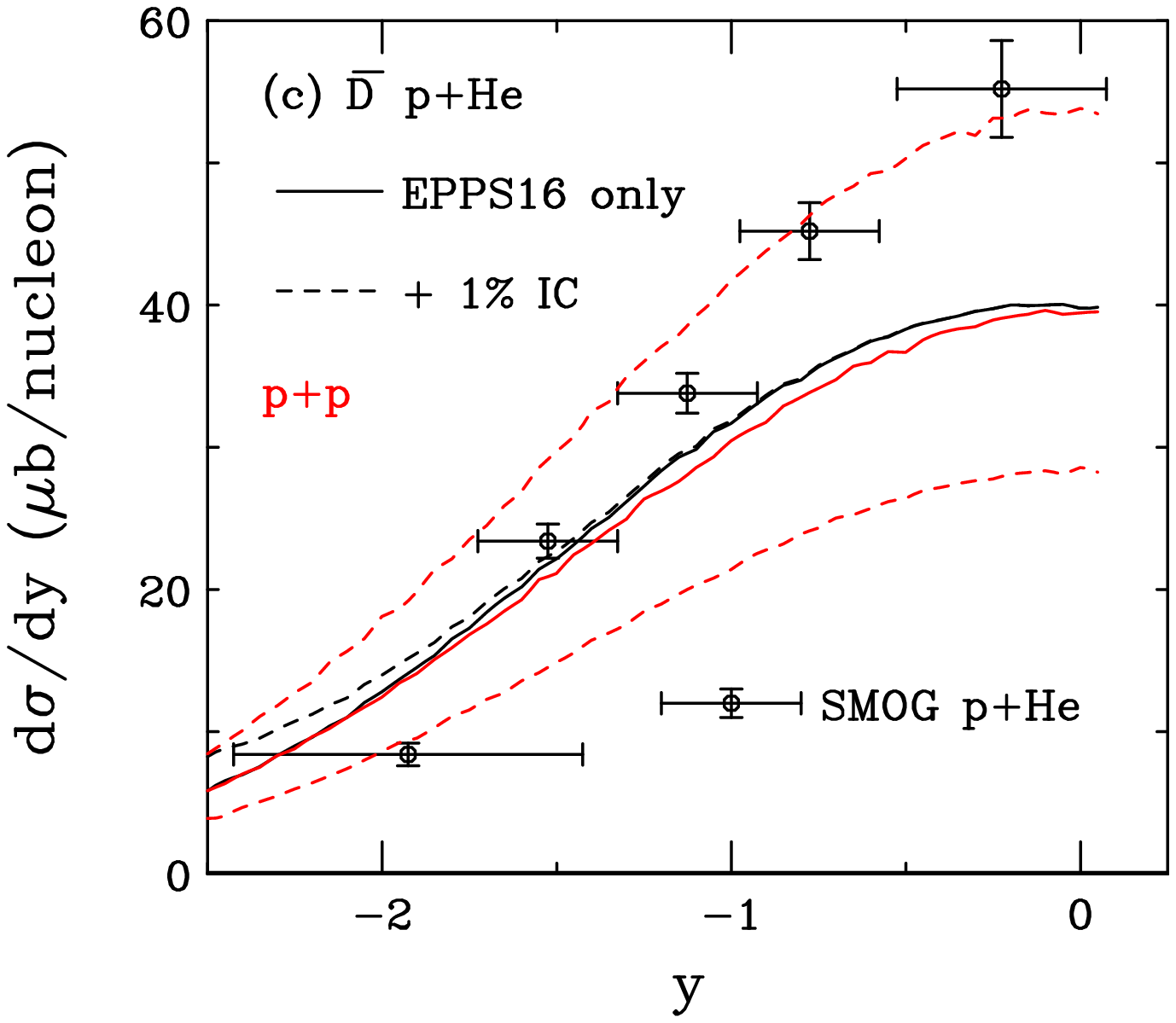}
    \includegraphics[width=0.4\textwidth]{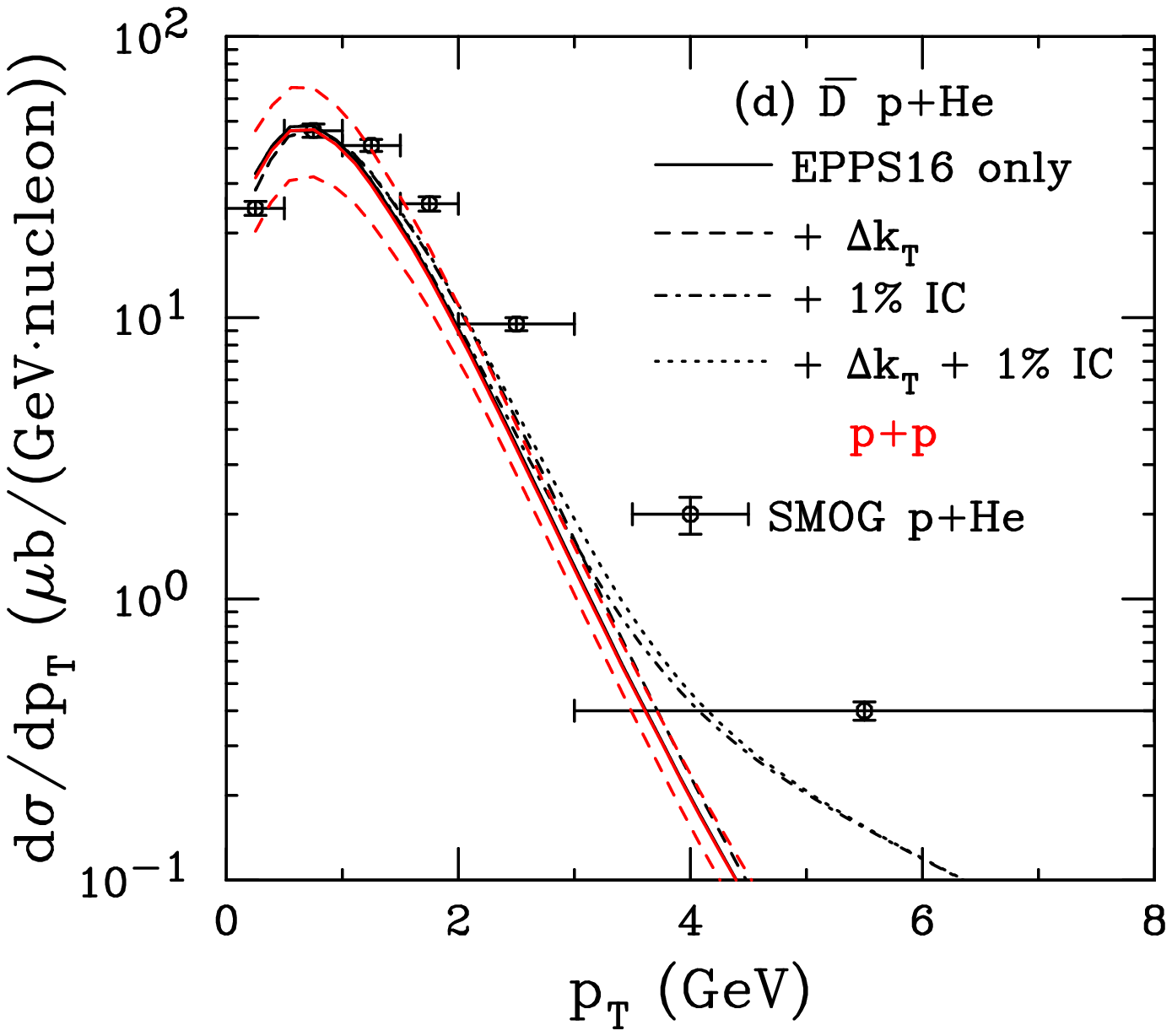} \\
    \includegraphics[width=0.4\textwidth]{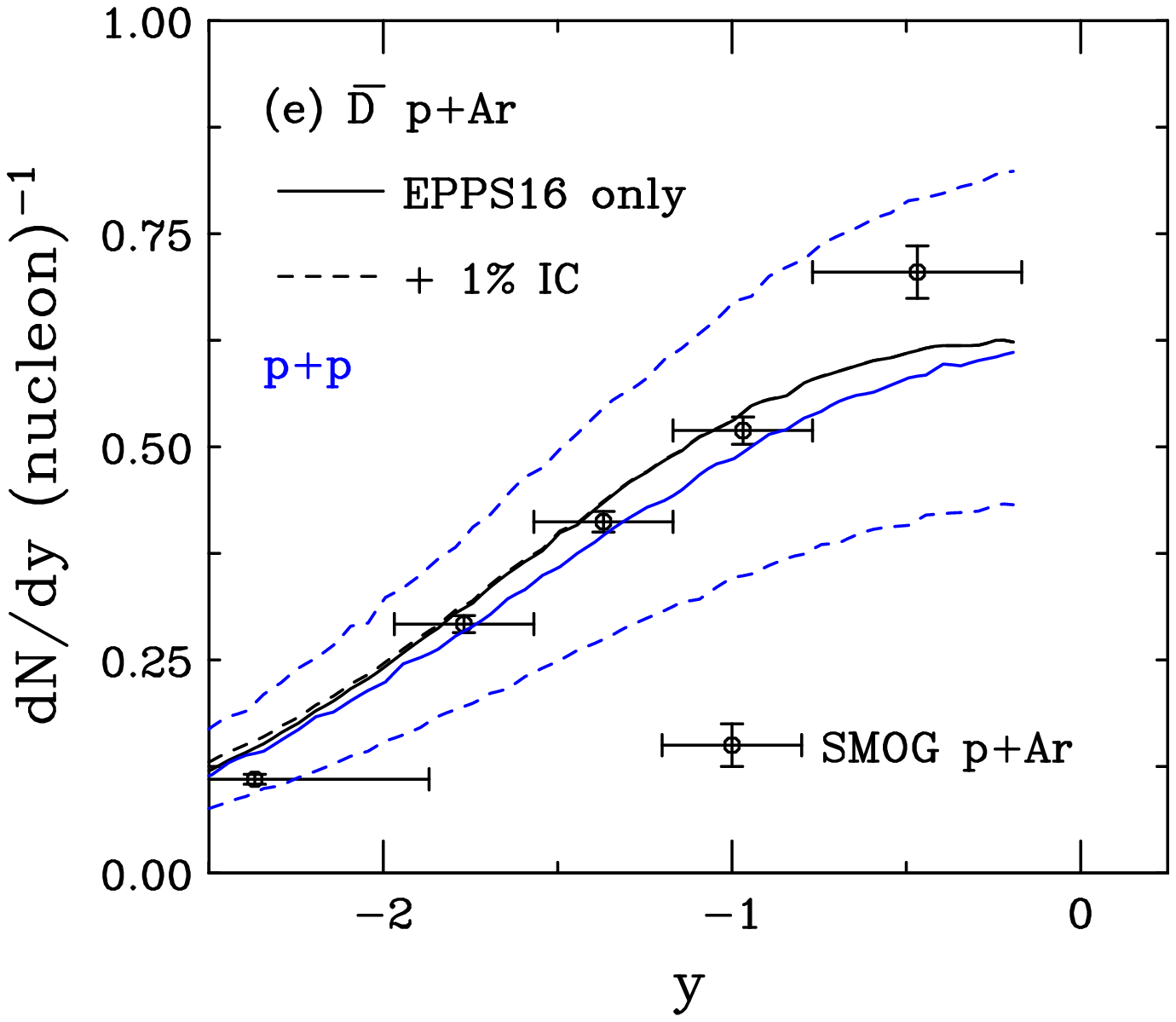}
    \includegraphics[width=0.4\textwidth]{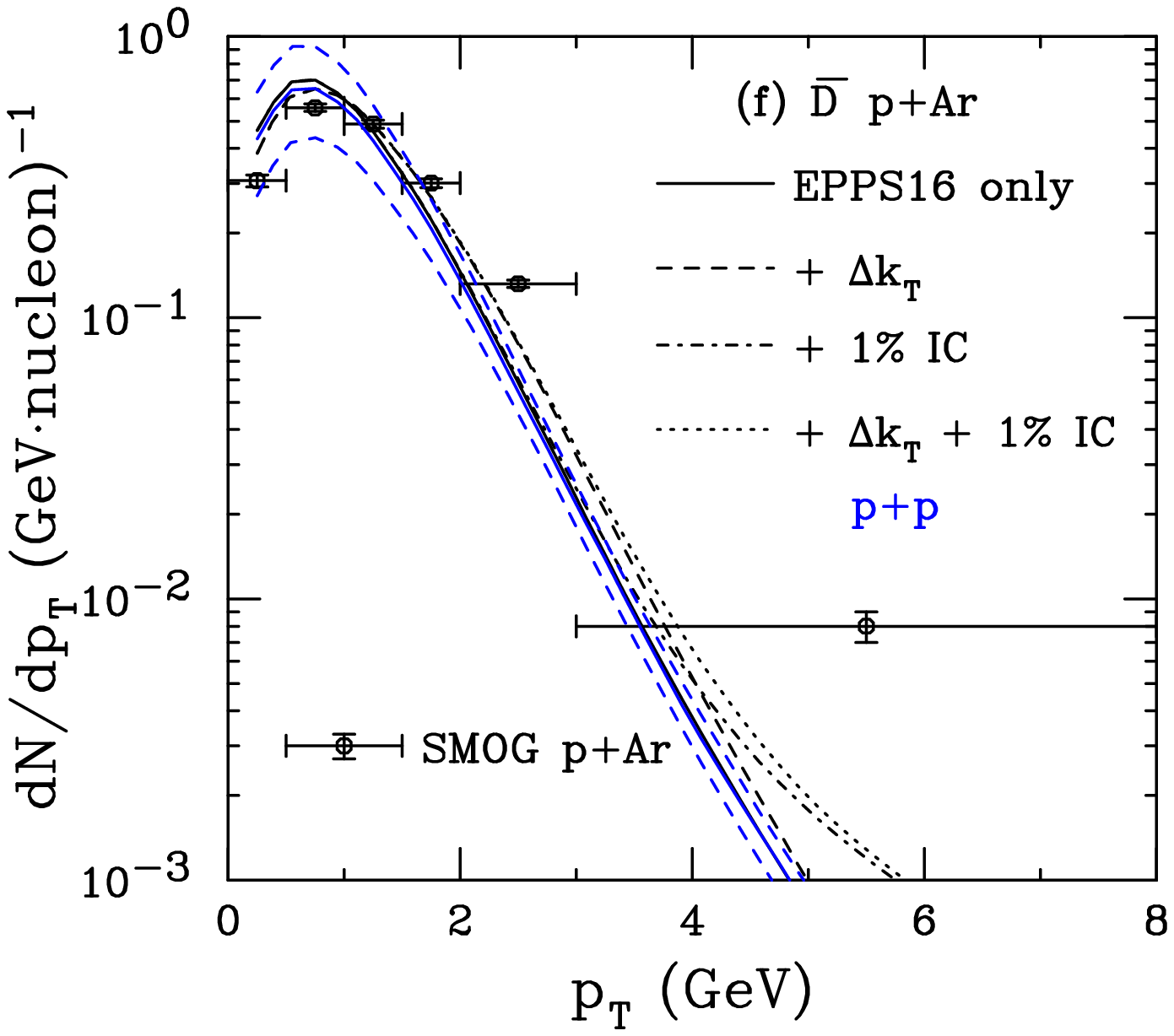}
  \end{center}
  \caption[]{The $\overline D$ cross section as a function
    of rapidity in (a), (c), (e) and transverse momentum in (b), (d), (f)
    for $p+{\rm Ne}$ collisions at $\sqrt{s_{NN}} = 68.5$~GeV in (a) and (b);
    $p+{\rm He}$ collisions at $\sqrt{s_{NN}} = 86.6$~GeV in (c) and (d); and
    $p+{\rm Ar}$ collisions at $\sqrt{s_{NN}} = 110.4$~GeV in (e) and (f).
    The black curves are the $p+A$ calculations.  The colored curves (solid and
    dashed) show the NLO $p+p$ calculations (no IC) at the same energy for the central
    value and the limits of the uncertainty band.  
    The $p+A$ rapidity distributions are shown for EPPS16 only (solid) and
    EPPS16 with $P_{\rm ic \, 5}^0 = 1$\% (dashed).  The $p_T$
    distributions show EPPS16 only (solid); EPPS16 with $k_T$ kick (dashed);
    EPPS16 and $P_{\rm ic \, 5}^0 = 1$\% (dot-dashed); and EPPS16,
    $k_T$ kick and $P_{\rm ic \, 5}^0 = 1$\% (dotted).  The SMOG $p+{\rm Ne}$
    data are from Ref.~\cite{SMOGpNeD} while the $p+{\rm He}$ and $p+{\rm Ar}$
    data are from Ref.~\cite{SMOG}.
    }
\label{fig:DpAdists}
\end{figure}

Figures~\ref{fig:JpAdists} and \ref{fig:DpAdists} present the $y$ and $p_T$ distributions 
in the SMOG acceptance.  In both cases, the results are presented with increasing energy: 
panels (a) and (b) are for $p+{\rm Ne}$ collisions at $\sqrt{s_{NN}} = 68.5$~GeV; (c) 
and (d) are for $p+{\rm He}$ collisions at $\sqrt{s_{NN}} = 86.6$~GeV; and, finally, 
(e) and (f) show the results for $p+{\rm Ar}$ collisions at $\sqrt{s_{NN}} = 110.4$~GeV.  
At each energy, the $p+p$ distributions, along with their uncertainties, reproduced from F
igs.~\ref{cem_pp} and \ref{d0_pp}, are also shown.  It is worth noting that while the 
center of mass energy increases from top to bottom in the figures, the target mass does 
not follow the same trend: $A = 20$ in (a) and (b); $A=4$ in (c) and (d); and $A = 40$ in (e) and (f).

The $p+A$ cross sections are given for several different scenarios.  While the nPDF 
uncertainties have been calculated for all of the $p+A$ scenarios, only the central 
values are shown to avoid cluttering the figures.  The uncertainties due to the nPDFs 
lie well within those of the $p+p$ cross sections.  The ratios of the $p+A$ to $p+p$ c
ross sections are not presented because LHCb did not so far measure the $p+p$ cross 
sections at the same energies and so do not present the data in terms of a nuclear
modification factor.  

The $J/\psi$ $y$ and $p_T$ distributions are presented in Fig.~\ref{fig:JpAdists}.  
The rapidity distributions are discussed first.  The solid
black curves show the $p+A$ result with EPPS16 nPDF effects only.  These results are 
generally compatible with the $p+p$ cross section at the most negative rapidity where 
the modifications are small but are enhanced relative to the $p+p$ cross section at 
$y \sim 0$ where the momentum fraction, $x$, probed is in the antishadowing region.  
The smallest enhancement is for $p+{\rm He}$ collisions, with the lowest $A$, while the 
largest is for $p+{\rm Ar}$ collisions with the greatest
value of $A$.
The black dashed curve includes both the EPPS16 modifications and nucleon absorption.  
When absorption is included, suppression of the cross section relative to the central 
$p+p$ cross section is observed at all energies.  The dot-dashed and dotted curves show 
the addition of IC to the cross section.  IC does not change the distribution at 
midrapidity but introduces a small enhancement for $y < -1$.  The IC contribution tends 
to decrease with energy because the distribution is pushed to more negative rapidity 
with increasing energy \cite{RV_IC_EN}.  Thus the effect should be smaller at 
$\sqrt{s_{NN}} = 110.4$~GeV than at 68.5~GeV.  On the other hand, IC is more suppressed 
for larger mass targets so that the reduction is smallest for the He target.  
Nonetheless, the overall contribution from IC is small at SMOG energies and to 
distinguish the presence of IC, high statistics data in smaller rapidity bins at the 
most negative rapidity are required.

Four $p+A$ calculations are shown for the $p_T$ dependence, also accompanied by the 
uncertainty band for $p+p$ collisions including perturbative production alone.  The 
solid curve is again the nPDF effects alone with the central value of EPPS16.  A small 
enhancement due to antishadowing is observed for $p+A$ relative to $p+p$, particularly 
at low $p_T$.  The 
smallest nPDF effect is for $p+{\rm He}$ while the largest is for $p+{\rm Ar}$.  The 
dashed curve includes enhanced $k_T$ broadening due to the passage of the proton through 
the nucleus.  The effect reduces the maximum of the distribution at low $p_T$ but makes 
the cross section softer at higher $p_T$.  As is the case for nPDF effects, the smallest 
additional broadening is for the lowest mass target, thus it is smallest for $p+{\rm He}$ 
and largest for $p+{\rm Ar}$.  In addition, the baseline $p+p$ intrinsic $k_T$ broadening 
increases slightly with $\sqrt{s_{NN}}$.  When absorption is included, the $p+A$ 
calculation shifts downward by a constant factor representative of the survival 
probability for that value of $A$.

Finally, IC is included.  As previously mentioned, the IC contribution at midrapidity 
decreases with increasing energy so that the rapidity-integrated $p_T$ distribution is 
suppressed at low $p_T$, see Ref.~\cite{RV_IC_EN}.  This suppression increases with 
increasing $\sqrt{s_{NN}}$.  However, at higher $p_T$, as the cross section calculated 
perturbatively begins to fall off more steeply, the long tail of the IC $p_T$ distribution 
appears, typically for $p_T > 4$~GeV.  This long tail is because the IC contribution to 
$J/\psi$ production can take almost all of the proton momentum, especially near 
midrapidity.  If the sum of the momenta carried by the $c$ and $\overline c$ approaches 
the total proton momentum, the slope of the distribution will change and eventually reach 
an energy-dependent endpoint \cite{RV_IC_EN}.  However, in the $p_T$ range covered by 
SMOG, especially with the rapidity acceptance including $y \sim 0$, this energy 
constraint is not reached.  Detection of an enhanced $J/\psi$ distribution at high $p_T$
 would be a clear signature of IC.  However, high statistics data at large $p_T$ are 
 needed, with widths no larger than 1~GeV for $p_T > 4$~GeV.

The overall agreement with the SMOG data is generally quite good.  The calculated 
central $p+A$ values match the data well.  The data also lie within the mass and scale 
uncertainties of the $p+p$ cross section.  If all the uncertainties (mass, scale, and 
nPDF) are added in quadrature, the resulting uncertainty band would be wider still.  
No uncertainties have been estimated or included on $\Delta k_T^2$ or $\sigma_{\rm abs}$.

Results for $\overline D$ production are shown in Fig.~\ref{fig:DpAdists}.  Only two 
$p+A$ curves are shown for the rapidity distribution, nPDF effects alone (solid curve) 
and including a 1\% IC contribution.  Antishadowing effects at midrapidity are apparent 
when compared to the $p+p$ results.  The effect of IC is again small, as expected.  
While the calculated rapidity distributions generally agree with the data, the 
calculations at the most central rapidities underestimate the SMOG data.  The results are, 
however, still consistent with the mass and scale uncertainties on the $p+pT$ cross 
section, an underestimate of the total uncertainty on the $p+A$ cross section, as 
previously discussed.

Four curves are again shown for the $\overline D$ $p_T$ distribution.  The solid and 
dashed black curves include only perturbative cold nuclear matter effects, showing an 
enhancement due to antishowing at low $p_T$ for the nPDF effects alone and a reduction 
at low $p_T$ with a harder cross section at high $p_T$.  The dot-dashed and dotted curves 
include a 1\% IC contribution to the nPDF only and nPDF plus enhanced $k_T$ broadening 
respectively.  The calculations without and with IC are both softer than the same 
distributions for $J/\psi$ shown in the previous figure.    This is because the 
$\overline D$ meson carries only a single charm quark in both cases.  When IC is 
included, the high $p_T$ enhancement due to this effect is also reduced.  The agreement 
with the SMOG data again, is generally good.  No conclusion can be drawn about the 
highest $p_T$ data point because the bin width is so large.  One might expect, however, 
that a small bin width would lie closer to the curves because the bulk of the cross 
section is at the lowest $p_T$ due to the steeply falling distribution.

\begin{figure}
  \begin{center}
    \includegraphics[width=0.4\textwidth]{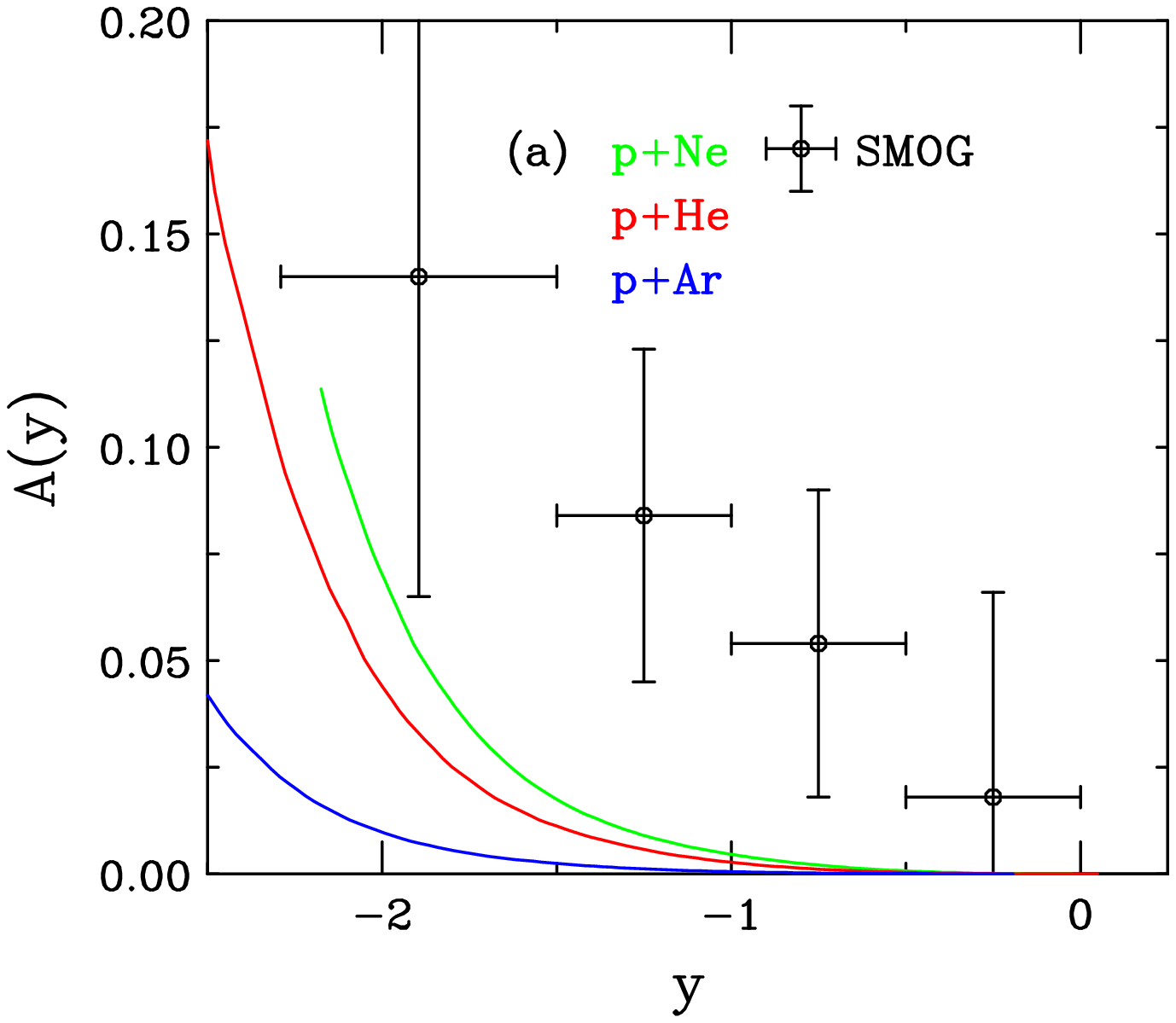}
    \includegraphics[width=0.4\textwidth]{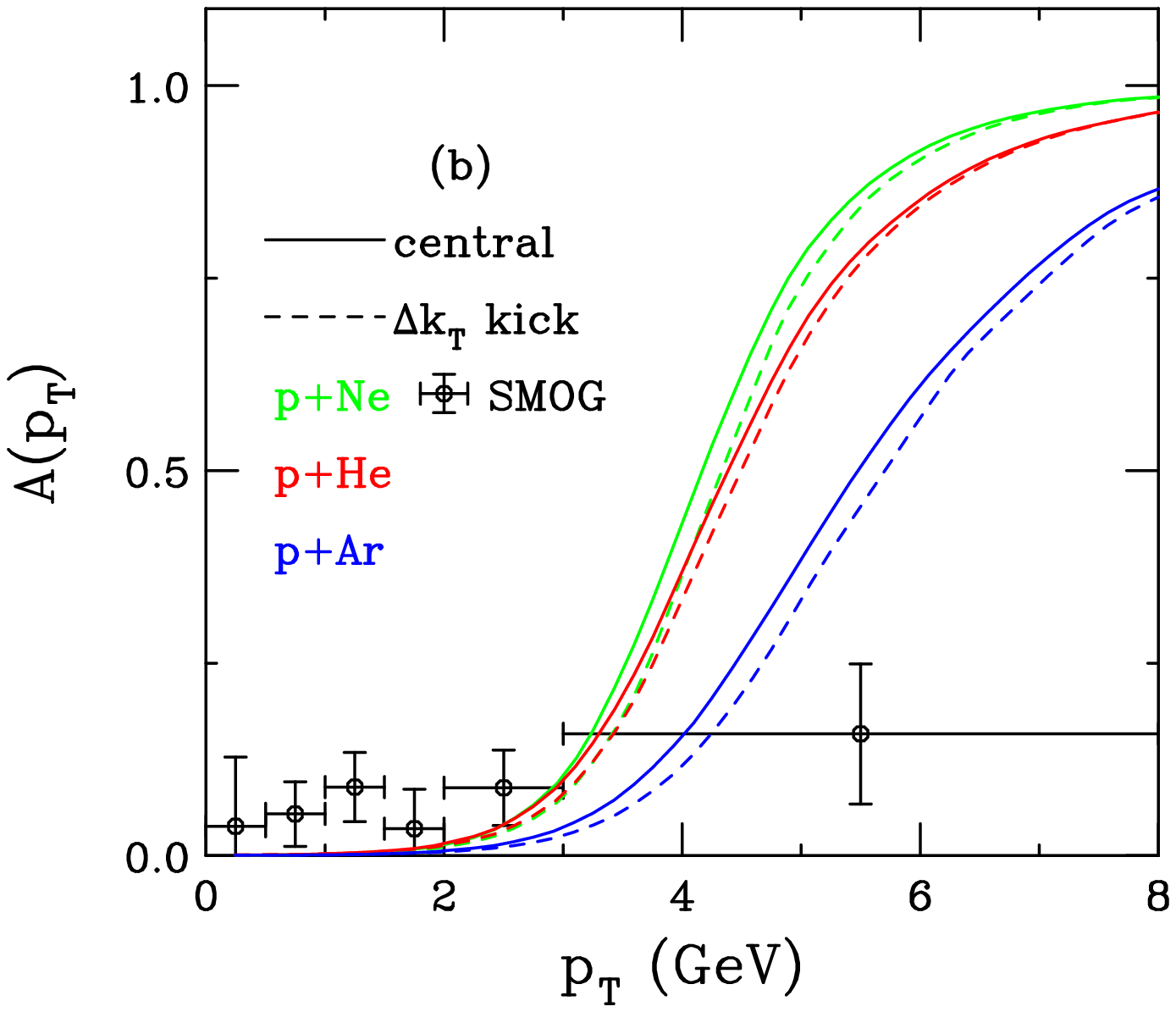}
  \end{center}
  \caption[]{The $\overline D$-$D$ asymmetry as a function
    of rapidity (a) and transverse momentum (b)
    for $p+{\rm Ne}$ collisions at $\sqrt{s_{NN}} = 68.5$~GeV (green);
    $p+{\rm He}$ collisions at $\sqrt{s_{NN}} = 86.6$~GeV (red); and
    $p+{\rm Ar}$ collisions at $\sqrt{s_{NN}} = 110.4$~GeV (blue).  All
    calculations include the EPPS16 central set and $P_{\rm ic \, 5}^0 = 1$\%.
    In (b) the dashed curves also include a $k_T$ kick.  The data are from
    Ref.~\cite{SMOGpNeD}.
    }
\label{fig:DpAasym}
\end{figure}

Because the effects of IC are small on the scale of individual distributions, especially 
when the effect is larger at the edge of the kinematic coverage, it is worth looking at 
the asymmetries between distributions. Forming the asymmetry between two distributions 
shows the effect on a linear scale, enhancing it.  There is no asymmetry for $J/\psi$ 
since both the $c$ and $\overline c$ are involved but there is one between 
$\overline D^0 (\overline c u)$ and $D^0 (c \overline u)$.  There would be a similar 
asymmetry between $D^-(\overline c d)$ and $D^+ (c \overline d)$.  As previously
discussed, the $\overline D^0$ and $D^-$ are leading charm mesons in a proton projectile 
because their valence quarks are shared with the valence quarks of the proton.  In the 
context of this work, these mesons can be produced directly from the lowest 5-particle 
IC state of the proton. The $D^0$ and $D^+$ are non-leading $D$ mesons because they would 
need to arise from a higher particle proton state, such as a 7-particle state and, in 
this case, the ``leading" and ``non-leading" $D$ distributions, as defined from the 
5-particle IC state would be identical.  Rather than increasing a potential asymmetry, 
including these states would tend to dilute it, as already mentioned in Sec.~\ref{IC}.

Such asymmetries have been observed in collisions involving a $\pi^- (\overline u d)$ 
projectile \cite{E769,E791,WA82} where now $D^0 (c \overline u)$ is leading relative 
to $\overline D^0 (\overline c u)$.  (Note that $D^- (\overline c d)$ is also now leading 
relative to $D^+ (c \overline d)$.)  These asymmetries have not previously been clearly 
observed in collisions with a proton projectile, potentially because the $\Lambda_c(udc)$ 
is the natural partner to $\overline D^0$ production from the 5-particle proton IC state.

The asymmetry between $\overline D^0$ and $D^0$ production is written as
\be
A_D(X) = \frac{(d\sigma_{\overline D^0}/dX) - (d\sigma_{D^0}/dX)}{(d\sigma_{\overline D^0}/dX) + (d\sigma_{D^0}/dX)} \, \, .
\ee
where $X = y$ or $p_T$.  The calculated asymmetry is shown in Fig.~\ref{fig:DpAasym}, 
both as a function of $y$ (a) and $p_T$ (b).  No uncertainties are shown because the 
perturbative cold nuclear matter effects, as well as the mass and scale uncertainties, 
cancel in the ratios as a function of rapidity.  There is a slight difference as a 
function of $p_T$, depending on whether or not enhanced $k_T$ broadening is included.  
(Note that here the asymmetry is defined as the diffreence between the leading and 
non-leading $D^0$ mesons while in the recent SMOG paper \cite{SMOGpNeD}, the asymmetry is 
defined as between the $c$ and $\overline c$, giving it the opposite sign in the 
publication relative to this work.)

The calculated asymmetry is negligible at midrapidity and increases toward backward 
rapidity.  It is largest for $p+{\rm Ne}$ collisions at the lowest energy.  The increase 
in the asymmetry is pushed back to more negative rapidity as the energy increases because 
the higher energy boosts the IC distribution to more backward rapidity, see 
Ref.~\cite{RV_IC_EN}.  The asymmetry in the measured range is not large, $A(y) < 0.2$, 
in all cases because the peak of the IC distribution in rapidity is at still more negative 
rapidity in the SMOG energy domain and the difference between the rapidity distributions 
with and without IC is small.

The measured asymmetry is generally larger and is finite near midrapidity.  While the 
measurement follows the same trend as the calculation, the calculation, assuming that the 
asymmetry arises from IC alone, underestimates the measurement.  No asymmetry is assumed 
to arise from the perturbative calculation itself.

There is a production asymmetry that can arise at next-to-leading order in perturbative 
QCD from the interference between $q+g$ and $\overline q + g$ contributions to heavy quark 
production relative to $q +\overline q$ production but this is small \cite{Nason:1989zy}.  
As shown in Ref.~\cite{Nason:1989zy}, $\overline c$ production was enhanced over $c$ 
production in $\pi^- N$ interactions at $\sqrt{s} = 23$~GeV, at high $x_F$, up to 15\% 
at $x_F = 0.8$, but negliglbe at $x_f \sim 0$.  A rapidity of $-2$ for 
$\sqrt{s} = 68.5$~GeV is equivalent to $x_F = -0.32$.  While taking these differences 
into account in $p+A$ collisions here would increase the asymmetry rather than wash it 
out, it is very small compared to that of IC alone and would not help explain the 
discrepancy.  Reference \cite{Nason:1989zy} noted that the $\overline c -c$ asymmetry 
arising from the intereference effects is separate from the previously measured leading 
vs. non-leading charm asymmetry \cite{E769,E791,WA82}.

On the other hand, the calculated asymmetry as a function of $p_T$ increases quickly 
above a few GeV.  At high $p_T$, where the IC distribution becomes harder than the perturbative 
distribution, $A(p_T) \rightarrow 1$, especially for $p+{\rm Ne}$ 
collisions at $\sqrt{s_{NN}} = 68.5$~GeV.  In this case, the asymmetry effectively 
saturates.  At the higher energies, the reduction due to the further boost of the rapidity 
distribution to more negative rapidity delays the rise in $A(p_T)$, particularly in 
$p+{\rm Ar}$ collisions.  Including broadening, indicated by the dashed line denoted 
`band limits' in the figure legend, slightly delays the overall increase in $A(p_T)$ in 
all cases because the perturbative $p_T$ distribution is hardened.  This effect is 
smallest for $p+{\rm He}$ collisions where $A=4$.  Note that while it might be possible 
to separate the rise in $A(p_T)$ at different energies and target masses, especially at 
the largest $A$ and highest energy, it is likely not possible to detect effects of 
enhanced $k_T$ broadening at high $p_T$ in the asymmetry.

Likewise, the $p_T$ asymmetry is underestimated at low $p_T$.  Even if there was an 
enhancement of $\overline c$ over $c$ at high $x_F$ or rapidity in $p+A$ collisions, 
the $p_T$ distribution, integrated over the longitudinal acceptance, would not be 
affected.  The highest $p_T$ data are relatively consistent with the calculations but this 
should be checked with higher statistics data and smaller bins at high $p_T$.  
Measurements at lower center of mass energies by a high intensity fixed-target experiment, 
such as NA60+ \cite{NA60p}, would produce an enhancement from IC at lower $p_T$ 
\cite{RV_IC_EN} and would provide a larger lever arm in energy for IC studies.

In Ref.~\cite{SMOGpNeD}, the asymmetry data at $\sqrt{s_{NN}} = 68.5$~GeV
was also compared to calculations combining a 1\% IC component with 10\%
recombination or coalescence which would favor $\overline D^0$ production
\cite{MS1,MS2}.
While this calculation gives a larger asymmetry at more negative
rapidity than that shown here, they find a $p_T$ asymmetry close to zero for
$p_T > 4$~GeV.

It would be interesting to test a potential coalescence effect by looking for an 
asymmetry between $\overline D^0$ and $D^-$ production.  In this case, if coalescence 
in the initial state were present, then for a proton, one would expect more 
$\overline D^0$ to be produced because there are two $u$ quarks in the proton and only 
one $d$ quark.  The SMOG targets are all noble gases with $Z \approx N$ and thus
with approximately the same number of $u$ and $d$ quarks in the targets with a slight 
preference for $d$ quarks in the Ar target.  It could also be interesting to check other 
targets where $N > Z$ to see if there are different asymmetries for non-isoscalar targets.

\section{Conclusions}
\label{conclusions}

The $J/\psi$ and $\overline D$ meson distributions have been compared to fixed-target 
SMOG data for $A = 4$, 20 and 40 and $\sqrt{s_{NN}} = 68.5$, 86.6, and 110.4~GeV.  
Cold nuclear matter effects on the perturbative calculation have been included, as well 
as intrinsic charm.  Agreement with the available SMOG data is very good overall, for 
both charmonium and open charm.

The results do not depend strongly on the nPDF set.  While only the
the $p+A$ results for the central EPPS16 set
are shown in Figs.~\ref{fig:JpAdists} and \ref{fig:DpAdists},
the variation in the
cross sections due to the nPDF sets is on the order of 20\%, smaller than the
uncertainties on the $p+p$ cross sections.  The effects of IC are small and do
not depend on the EPPS16 uncertainties.  Choosing another nPDF set would not
change this conclusion, especially in the region where IC is becoming important,
at the most negative rapidities and highest $p_T$,
because the nPDFs typically differ more at low $x$ than at high $x$ where IC can
be important.  (See Ref.~\cite{RV_LHCshad} for a discussion of nPDF
variations.)

The current SMOG data cannot clearly distinguish between 
the presence or absence of IC due to the boost of the effect to higher rapidity
at increased energy.  Distinguishing the effects of IC on $J/\psi$ in the SMOG
apparatus would be difficult as a function of rapidity because the effect is
small in the SMOG rapidity acceptance but could be measured if
SMOG could gather sufficient statistics at high $p_T$ to increase the number of
$p_T$ bins above 3~GeV.  This could be feasible with upgrades to SMOG and
increased luminosity.

Similar statements can be made with respect to $D$ meson measureemnts.
However, studying the asymmetry between leading and non-leading $D^0$ mesons  
could help amplify the effect of IC with higher statistics data although intrinsic charm
alone underestimates the measured SMOG asymmetry.

{\bf Acknowledgments}
I would like to thank E. Maurice and V. Cheung
for helpful discussions.  Increased post-pandemic air travel on United Airlines
gave me the time and opportunity to finish the paper.
%and United Airlines for quiet time to work on the text. 
This work was supported by the Office of Nuclear
Physics in the U.S. Department of Energy under Contract DE-AC52-07NA27344 and
the LLNL-LDRD Program under Projects No. 21-LW-034 and 23-LW-036 and by
the U.S. Department of Energy, Office of Science, Office of Nuclear Physics
through the Topical Collaboration in Nuclear Theory on Heavy-Flavor Theory
(HEFTY) for QCD Matter under award no. DE-SC0023547.

\end{document}